\documentclass[a4paper,12pt,oneside]{article}  
\usepackage{geometry}
\usepackage[english]{babel}
\usepackage{microtype}
\usepackage{natbib} 

\usepackage{import, multicol,lipsum} 
\usepackage[utf8]{inputenc} 
\usepackage[T1]{fontenc}
\usepackage{subcaption}
\usepackage{pifont}
\usepackage{xcolor, blindtext}
\usepackage{url}
\usepackage{hyperref} 
\hypersetup{
colorlinks=true, 
linkcolor=red,
citecolor=blue,     
urlcolor=blue
}

\usepackage{emptypage} 
\usepackage[export]{adjustbox} 
\usepackage{amsthm, amsmath, amssymb, amsfonts, mathtools} 
\usepackage{nicefrac}
\usepackage{enumitem}
\usepackage{graphicx} 
\usepackage{multirow, nicematrix, array} 
\graphicspath{{GNHAR-img/}} 
\usepackage{wrapfig}
\usepackage{float} 
\usepackage{tikz} 
\usepackage{verbatim}
\usepackage{bbm}
\usepackage[capitalise]{cleveref}
\RequirePackage{fix-cm}
\usepackage{booktabs}

\newtheorem{theorem}{Theorem}
\newtheorem{definition}[theorem]{Definition} 
\theoremstyle{remark}
\newtheorem*{remark}{Remark}

\newcommand{\set}[1]{\left\{#1\right\}}

\def \R {\mathbb{R}}
\def \Z {\mathbb{Z}}
\def \Nat {\mathbb{N}}
\def \E {\mathcal{E}}
\def \G {\mathcal{G}}

\def \N {\mathcal{N}}
\def \V {\mathcal{V}}

\def \mmu {\boldsymbol{\mu}}
\def \vveps {\boldsymbol{\varepsilon}}

\def \LC {\mathbf{LC}}
\def \LJ {\mathbf{LJ}}
\def \XX {\mathbf{X}}

\def \YY {\mathbf{Y}}

\title{
    \bf{Network Time Series Models for \\ Multivariate Volatility Forecasting}
}

\author{
    \begin{tabular}{c}
        Chiara Boetti \\ \texttt{cb2605@bath.ac.uk}
    \end{tabular}
    \hspace{2cm}
    \begin{tabular}{c}
        Matthew A. Nunes \\ \texttt{man54@bath.ac.uk}
    \end{tabular} \\ [2ex] 
    Department of Mathematical Sciences, University of Bath
}

\date{}

\begin{document}
\maketitle
\thispagestyle{empty}

\begin{abstract}
\footnotesize
Realized volatility has become a standard tool for measuring latent variation in financial assets, and its forecasting is crucial for a wide range of financial applications.
We propose a network-based model for forecasting a vector of realized variance processes through the heterogeneous autoregressive (HAR) approach. The \textit{generalised network HAR} (GNHAR) model incorporates cross-sectional spillovers through a directed graph inferred from Granger-causality tests or connectedness indices, yielding a parsimonious multivariate time series  model specification. In an application to ten equities over tranquil and crisis regimes, the proposed GNHAR model improves upon common HAR model benchmarks under both short- and long-term forecasting. We also compare the network-based specification when the jump-continuous decomposition or node-specific option-implied variances are considered. Finally, unlike overparameterised models, our approach yields a concise set of parameters that track the strengthening or weakening of cross-market dependencies, providing a time-varying quantitative assessment of market stability.
\end{abstract}


\section{Introduction}
Accurate forecasting of volatility, i.e., the time-varying variability in a financial asset's price, is essential for risk management and for designing optimal decision-making policies. In practice, volatility is an unobservable process, as only prices or returns can be observed. Early work to model such data relied on daily (or other low-frequency) returns and inferred volatility using parametric models, such as generalised autoregressive conditional heteroskedasticity (GARCH) models or continuous-time stochastic-volatility models \citep{engle1982autoregressive, bollerslev1986generalized, taylor1994modeling}. See \citet{poon2003forecasting} for a review of volatility forecasting.

Subsequently, the widespread availability of high-frequency data, such as intraday returns recorded every 5 or 10 minutes, has enabled the development of realized measures, namely non-parametric estimators based on these intraday returns \citep{barndorff2002econometric, andersen2003modeling}. These realized measures make volatility empirically measurable, providing not only new ways to complement latent-variable approaches, but also motivating the development of models that work directly with such realized measures.
For instance, realized variance (RV) computed from 5-minute intraday returns is widely used to study volatility dynamics, as it offers a practical trade-off between information content and microstructure noise \citep{bandi2006separating, hansen2006realized}.
 
In this context, \citet{corsi2009simple} developed the heterogeneous autoregressive (HAR) model for realized volatility, which models volatility dynamics through a parsimonious combination of daily, weekly, and monthly components. Thanks to its simplicity and strong empirical performance, the HAR model has become a standard benchmark for forecasting RV.
In recent years, many extensions of the HAR model have been explored. For instance, \citet{andersen2007roughing} propose modelling the jump and continuous components of RV separately. Other extensions investigate the role of exogenous predictors under different settings. For example, \citet{liu2015economic} and \citet{audrino2020impact} study how economic and sentiment variables affect volatility forecasts across industries and market capitalisations, while \citet{shi2020comparison} and \citet{liang2020implied} compare different market conditions and assesses their influence on the conditional predictive ability of realized measures and option-implied volatilities. 

In the multivariate setting, modelling and forecasting volatility has taken two main directions. The first focuses on the realized covariance matrix processes \citep{chiriac2011modelling}. In particular, using the variance-correlation (DRD) decomposition, one can model the dynamics of the realized covariance by focusing on its realized variances and realized correlations separately. This is known as the HAR-DRD approach \citep{oh2016high}. More recently, \citet{zhang2025graph} and \citet{tapia2025higher} improved HAR-DRD by introducing graph-based methods to better capture cross-sectional dependencies among assets. In these models, each asset is represented as a node in a graph, and edges indicate spillover effects between volatilities.  Similarly, \citet{zhang2025forecasting} developed a graph neural network model based on a undirected graphical lasso estimated network to capture non-linear relations. Within an exogenous framework, \citet{luo2025forecasting} analyses HAR-DRD models with exogenous variables spanning financial, macroeconomic, sentiment, and climate indicators. 

The second direction extends the HAR framework by focusing on vectors of RV processes directly \citep{bubak2011volatility}, using vector HAR (VHAR) model.
The benefit of modelling vectors of RV as a multivariate process, as opposed to modelling each RV separately with simple HAR models have been recently evaluated by \citet{wilms2021multivariate}. More precisely, the work compared univariate and vector HAR models across several settings, including whether to use the jump-continuous decomposition and whether to include option-implied variance as an exogenous component, concluding that VHAR models can provide gains compared to univariate HAR models, especially at longer horizons.

Beyond forecasting accuracy, these vector models can also provide insight into market interconnections and the degree of co-movement between processes. For example, by comparing estimated coefficients, one can assess the strength of relationships among stock volatilities. Several extensions of VHAR models have been proposed to capture different types of dependence in financial markets. For instance, \citet{taylor2015realized} imposes cross-sectional dependence restrictions when modelling the RV of international stock markets. Similarly, \citet{cubadda2017vector} proposes a restricted VHAR model in which each horizon affects outcomes only through common indices, and finds it yields more accurate forecasts than a diagonal VHAR specification, which no interactions among stocks are assumed. In practice, results from VHAR models are often compared with model-agnostic measures to identify cross-market spillovers. Two widely used tools are Granger causality tests \citep{granger1969investigating}, which assess predictive causal links among time series, and the Diebold-Yilmaz connectedness index \citep{diebold2012better}, which captures both direct and indirect propagation of shocks through the system. Information from Granger causality tests or the Diebold-Yilmaz connectedness index can be represented as a network, where edges indicate spillover effects between assets.

It is less common, however, to explicitly incorporate prior information on spillovers when forecasting vector RV processes. For example, \citet{son2023forecasting} first identify relevant links between markets using either the Pearson correlation or the Diebold-Yilmaz connectedness index, subsequently incorporating these links into a non-linear graph neural network, demonstrating the forecasting benefits of capturing non-linear interactions among the components.
Motivated by these observations, we further investigate the role of graph-based modelling in forecasting multivariate RV processes, specifically employing linear models. Network-based linear models not only explicitly account for asset interconnections, in contrast to the baseline HAR framework, but they also bypass the overfitting issues common in non-linear models. Furthermore, thanks to their linear specification, they yield interpretable coefficients, allowing for a deeper understanding of the mechanisms governing volatility dynamic.

In this article, we propose an alternative approach to current multivariate RV models that explicitly incorporates a network of stock spillovers to forecast a vector of RV processes. Specifically, we exploit the \textit{generalised network HAR} (GNHAR) model, which combines recently proposed network time series models \citep{knight2016arxiv, zhu2017network, knight2020generalized} with the HAR modelling approach \citep{corsi2009simple}. While the GNHAR framework has previously been explored in the HAR-DRD context \citep{zhang2025forecasting, tapia2025higher}, to our knowledge it has not yet been used to model the vector of RV processes directly.
Given a fully-connected graph, or a directed graph inferred from Granger causality tests or Diebold-Yilmaz connectedness index, the proposed model outperforms univariate benchmarks in forecasting across various horizons. We also introduce two novel analyses to the existing literature: investigating how jump-continuous decomposition affects prediction accuracy within different graphical configurations, and examining the role of option-implied variance as a node-specific exogenous process across various degrees of time and network dependence.
Moreover, our approach provides insight into the magnitude of individual stock dynamics and their interactions through the estimated model coefficients, helping to clarify the role of the hierarchical HAR components during both tranquil periods and financial crises.

The remainder of this article is organised as follows. \cref{sec:background} introduces notation and background regarding the HAR model and network time series models. \cref{sec:methodology} presents our proposed GNHAR model and its jump-continuous extensions. After describing the RV dataset of ten assets and the experimental design in \cref{sec:data}, we compare the forecasting performance of the various modelling approaches in \cref{sec:forecast}. We analyse the option-implied variance as exogenous regressor within the GNHAR framework in \cref{sec:exogenous}, and inspect the various model components through their estimated parameters. Concluding remarks are provided in \cref{sec:concs}.

\section{Background Material}\label{sec:background}
In this section, we outline the pertinent background to motivate our proposed models in \cref{sec:methodology}.  More specifically, following a review of the original HAR model \citep{corsi2009simple} and its extensions, we provide an overview of network time series models \citep{knight2016arxiv, knight2020generalized}.

\subsection{The Heterogeneous Autoregressive Modelling Framework}
The heterogeneous autoregressive (HAR) model, introduced by \citet{corsi2009simple}, is based on a hierarchical multi-horizon representation. Volatility is modelled using components computed over different aggregation windows (e.g., daily, weekly, and monthly), reflecting heterogeneity in market participants and providing a flexible yet parsimonious forecasting framework.
\begin{definition}[Heterogeneous Autoregressive Model]
    Given a univariate realized variance (RV) process, $\set{RV_{t}}_{t\in\Z}$, the HAR model is defined as 
    \begin{equation}\label{eq:HAR}
  RV_t = \alpha^{(d)} RV^{(d)}_{t-1} + \alpha^{(w)} RV^{(w)}_{t-1} + \alpha^{(m)} RV^{(m)}_{t-1} + \varepsilon_t,
    \end{equation}
    where $\set{\varepsilon_t}_{t\in\Z}$ is the uncorrelated, zero mean and constant variance error term. Each $RV^{(\cdot)}_{t}$ is the aggregated RV over the last $k$ days, namely $RV^{(\cdot)}_{t} = \frac{1}{k} \sum_{j=0}^{k-1} RV_{t-j}$ with $k=1,5$, and $22$ for $RV^{(d)}_{t}$, $RV^{(w)}_{t}$, and $RV^{(m)}_{t}$, respectively.
\end{definition}
In model \eqref{eq:HAR}, the parameters $\alpha^{(d)}$, $\alpha^{(w)}$, and $\alpha^{(m)}$ account for the daily, weekly, and monthly aggregation, respectively. The daily coefficient captures the short-term behaviour, whereas the monthly coefficient accounts for the long-term dependencies.
In practice, the model is estimated on log-transformed RV to approximate normality and to mitigate the impact of large volatility spikes on the estimation. In the following, we denote the log-transformed RV process as $X_{t}:= \log(RV_{t})$. 

\paragraph{Jump-Continuous decomposition.}
\citet{andersen2007roughing} model the RV process by explicitly disentangling the continuous path from the jump dynamics. In particular, given the bi-power variation, $\set{BPV_{t}}_{t\in\Z}$, the jump component of a certain asset is defined as
\begin{equation}\label{eq:jump}
    J_{t} := \max\set{RV_{t}-BPV_{t}, 0},
\end{equation}
and the continuous component is $C_{t} := RV_{t} - J_{t}$, for each $t\in\Z$. 
When focusing on the log-transformed process, the JC-HAR model then becomes 
\begin{equation}\label{eq:JC-HAR}
\begin{aligned}
    X_{t} =
    \alpha_{J}^{(d)} LJ^{(d)}_{t-1} + \alpha_{J}^{(w)} LJ^{(w)}_{t-1} + \alpha_{J}^{(m)} LJ^{(m)}_{t-1} 
    + \alpha_{C}^{(d)} LC^{(d)}_{t-1} + \alpha_{C}^{(w)} LC^{(w)}_{t-1} + \alpha_{C}^{(m)} LC^{(m)}_{t-1} + \varepsilon_t,
\end{aligned}
\end{equation}
where $LJ_{t}^{(\cdot)} = \frac{1}{k} \sum_{j=0}^{k-1} \log(J_{t}+1)$ are the daily ($k=1$), weekly ($k=5$), and monthly ($k=22$) aggregated jump components, respectively. Similarly, $LC_{t}^{(\cdot)} = \frac{1}{k} \sum_{j=0}^{k-1} \log(C_{t})$ denotes the relative aggregated continuous component.

\paragraph{Inclusion of exogenous information.}
The HAR model in \eqref{eq:HAR} can be extended to account for additional information through an exogenous component. More precisely, the HAR model with exogenous process $\set{Y_{t}}_{t\in\Z}$ of autoregressive (AR) order $p\in\Nat$, HARX($p$), is defined as
\begin{equation*}
    X_{t} = \alpha^{(d)} X^{(d)}_{t-1} + \alpha^{(w)} X^{(w)}_{t-1} + \alpha^{(m)} X^{(m)}_{t-1} + \sum_{j=1}^{p} \lambda_{j} Y_{t-j} + \varepsilon_t.
\end{equation*}
Examples of exogenous processes used for forecasting RV include financial variables such as jump variation, option-implied volatility, and other realized measures \citep{andersen2007roughing, liang2020implied, wen2019forecasting}; macroeconomic indicators such as economic policy uncertainty \citep{liu2015economic}; or attention-, sentiment-, and climate-based predictors \citep{audrino2020impact, luo2025forecasting}. 

\subsection{Generalised Network Autoregressive Processes}
A network time series is a multivariate process defined over a \textit{network} or \textit{graph}, in which each node corresponds to a univariate time series, and a possible connection between two components is modelled via the edge of the given network. Hence, network time series models provide a flexible modelling framework allow to naturally account for the connections among the various time-dependent variables.

\paragraph{Graph Definitions.}
Let $\G = (\V, \E)$ be a graph, where $\V = \set{1,\ldots,N}$ denotes the set of vertices or nodes, and $\E$ is the set of edges. The graph is said to be a \textit{directed} graph if $\E = \set{(i,j)\in \V\times\V : i\rightsquigarrow j}$, where $i \rightsquigarrow j$ indicates that a connection exists from the $i$-th node to the $j$-th node. Alternatively, for an \textit{undirected} graph, we write $i\leftrightsquigarrow j$ to indicate a pairwise connection between the two nodes $i$ and $j$. We assume $\G$ has no self-loops, that is, edges of the form $(i,i)$ are not permitted. 

For each $i\in\V$, we define the \textit{$r$-stage neighbour set of $i$}, $\N^{(r)}(i)$, as 
\begin{equation*}
\left\{
\begin{aligned}
    \N^{(1)}(i) & = \set{j\in\V\setminus \mathcal{I} : i \rightsquigarrow j, i\in \mathcal{I}} \\
    \N^{(r)}(i) & = \N\set{\N^{(r-1)}(i)}\setminus \left[ \set{ \bigcup_{q=1}^{r-1} \N^{(q)}(i)} \cup \set{i} \right] \quad \text{for } r=2,3,\ldots  
\end{aligned}
\right.
\end{equation*}

Another relevant quantity in network analysis is the weighted adjacency matrix $W \in \R^{N,N}$, with entries $w_{i,j} \in [0,1]$ for all $i, j \in \V$.
For an unweighted graph $\G$, all connections are equally important and $w_{i,j} = 1/|\N^{(r)}(i)|$, with $|\cdot|$ denoting the cardinality of the set.
In the context of weighted networks, a raw weight $\tilde{w}_{i,j} \ge 0$ captures the association strength between two nodes $i, j \in \V$, and $\tilde{w}_{i,j} = 0$ if $(i,j) \notin \E$. In this case, those weights are normalised such that their sum within each $r$-stage neighbourhood is equal to one, i.e.,~$w_{i,j} = \tilde{w}_{i,j}/(\sum_{j \in N^{(r)}(i)} \tilde{w}_{i,j})$.
Finally, the \textit{$r$-stage weighted adjacency matrix} $W^{(r)} \in \R^{N,N}$ encodes the normalised weighted structure associated with the $r$-stage neighbours of $\G$, and its entries are defined as~$W^{(r)}_{i,j} = w_{i,j}  \mathbbm{1}\!\left(j\in\N^{(r)}(i) \right)$.

\paragraph{The GNAR Model.}
Given a network $\G=(\V,\E)$ and $\XX_{t} = \left(X_{1,t}, \ldots, X_{N,t} \right)^{\top}$ for each $t\in\Z$, we assign each $\set{X_{i,t}}_{t\in\Z}$ to a node $i\in\V$ and assume the edge $(i,j)\in\E$ indicates a connection with another component $\set{X_{j,t}}_{t\in\Z}$, for $j\neq i$. Hence, $\set{\XX_{t}}_{t\in\Z}$ is a time series process whose components are observed over the nodes of $\G$. \citet{knight2020generalized} introduce the generalised network autoregressive (GNAR) framework to model such multivariate processes.

\begin{definition}[Generalised Network Autoregressive Process]
    Let $\G=(\V,\E)$ be a graph, and let $\set{\vveps_{t}}_{t\in\Z}$ be a zero-mean Gaussian process, with covariance matrix $\Sigma_{\vveps} = \sigma^{2} I_{N}$. 
    Then, $\set{\XX_{t}}_{t\in\Z}$ is a GNAR process of order ($p,\boldsymbol{s}$), with $\boldsymbol{s}= (s_{1}, \ldots, s_{p})$, if, for all $t\in\Z$,
    \begin{equation}\label{eq:GNAR}
  \XX_{t} = A_{1} \XX_{t-1} + \ldots + A_{p} \XX_{t-p} + \vveps_{t},
    \end{equation} 
    and $A_{l} \in \R^{N \times N}$ are defined as
    \begin{equation*}
  A_{l} = \operatorname{diag}(\alpha_{1,l},\ldots,\alpha_{N,l}) + \sum_{r=1}^{s_{l}} \beta_{l,r} W^{(r)},
    \end{equation*}
    for $l=1,\ldots, p$. This class of processes is denoted GNAR($p,\boldsymbol{s}$).
\end{definition}

With this notation, $p \in \Nat$ is the maximum AR lag, and $s_{l} \in \Nat$ is the largest stage of neighbour network dependence at lag $l$, for $l=1\ldots, p$. Thus, AR components are characterised by their own sets of parameters: $\alpha_{i, l} \in \R$, the usual AR coefficient for node $i$ at lag $l$, and $\beta_{l, r} \in \R$, the network AR coefficient capturing the influence of the $r$-stage neighbours at lag $l$.

The model defined in \cref{eq:GNAR} is commonly referred to as \textit{individual}-$\alpha$ model. Alternatively, we refer to a \textit{global}-$\alpha$ model when a common AR parameter is shared by all nodes at each lag $l$, i.e.,~$\alpha_{i,l}=\alpha_{l}$ for each $i=1,\ldots, N$. \cref{eq:GNAR} can be also generalised to node-specific network AR parameters, leading to the \textit{local}-$\set{\alpha\beta}$ model \citep{nason2025forecasting}, or to \textit{community}-$\set{\alpha\beta}$ model, where the same coefficients are shared by the nodes within the same community \citep{nason2024modelling}.

While the GNAR($p,\boldsymbol{s}$) specification can be regarded as a constrained version of the VAR($p$) model, it offers several benefits, such as handling missing values more effectively. Since each variable pair corresponds to a nearest neighbour, all such pairs in the data contribute to the estimation of the nearest neighbour parameter \citep{knight2020generalized}. Thus, even if one node contains a large proportion of missing observations, the abundance of other pairs typically ensures that the parameter remains well-estimated. 

Another advantage lies in the higher-order interactions captured through the graph, while preserving a parsimonious parametrisation. Specifically, the network dependence parameters $\{\beta_{l,r}\}$ facilitates interactions among components differently according to both the neighbourhood stage and the lag. However, compared to a VAR($p$) model, where the number of parameters grows quadratically with the dimension of the process, $\mathcal{O}\left(pN^{2}\right)$, the number of parameters in the individual-$\alpha$ GNAR($p,\boldsymbol{s}$) model grows linearly, $\mathcal{O}\left(pN + \sum_{l=1}^{p} s_{l}\right)$. This is further reduced to a constant rate with the global-$\alpha$ specification, $\mathcal{O}\left(p + \sum_{l=1}^{p} s_{l}\right)$.
For these reasons, the GNAR framework and its extensions have proven efficient in modelling and forecasting in a variety of settings, including macroeconomic indicators \citep{nason2022quantifying, nason2025forecasting} and realized volatility measures \citep{boetti2025long, tapia2025higher}.

\section{Modelling Realized Variances via Network Time Series}\label{sec:methodology}
In modern financial markets, the dynamics of assets and indices are tightly coupled within a complex system and shocks to one asset can have a rapid effect on the other assets. Over recent years, the financial literature regarding the use of networks to describe connections among stock markets has grown substantially. Indeed, a graphical representation not only allows for an easier interpretation of cross-interactions, but it also provides a mathematically parsimonious way to capture intricate, high-dimensional spillover effects. As a consequence, this leads to stable and accurate methods for modelling and forecasting volatility.

In this section, we investigate a network-based approach for modelling multivariate RV process when such a financial graphical representation is known. In addition to capturing the daily, weekly, and monthly behaviour of individual stocks, this approach can exploit connections among stocks across different time horizons. By incorporating interactions at different time horizons, the network representation is flexible and informative, improving forecasting performance while providing insight into the volatility dynamics through a relatively small number of estimated parameters.

Throughout the rest of the article, we consider $N$ assets and we denote by $X_{i,t} := \log(RV_{i,t})$ the daily log-transformed RV of the $i$-th asset, computed from 5-minute intraday returns, for~$i=1,\ldots,N$ and $t\in\Z$. Thus, $\set{\XX_{t}}_{t\in\Z}$ is the collection of these $N$-dimensional vectors processes, with $\XX_{t} = \left(X_{1,t}, \ldots, X_{N,t} \right)^{\top}$ for each $t\in\Z$.

\subsection{Generalised Network Heterogeneous Autoregressive Model}
Our proposed \textit{generalised network heterogeneous autoregressive} (GNHAR) model combines the aggregated structure of \cref{eq:HAR} with the network-based model of \cref{eq:GNAR}. 
\begin{definition}[Generalised Network Heterogeneous Autoregressive Model]
    Given a graph $\G=(\V,\E)$, the GNHAR model of order ($s^{(d)},s^{(w)},s^{(m)}$) over $\G$ is defined as
    \begin{equation}\label{eq:GNHAR}
        \XX_{t} = \mmu + A^{(d)} \XX^{(d)}_{t-1} + A^{(w)} \XX^{(w)}_{t-1} + A^{(m)} \XX^{(m)}_{t-1} + \vveps_{t},
    \end{equation}
    for all $t \in \Z$. Here, $\mmu$ is the vector of intercept parameters, $\set{\vveps_t}_{t}$ is a zero-mean Gaussian process with covariance matrix $\sigma^{2} I_{N}$, and $\XX^{(d)}_{t}$, $\XX^{(w)}_{t}$, and $\XX^{(m)}_{t}$ are the daily, weekly, and monthly aggregated components with corresponding coefficient matrix
    \begin{equation}\label{eq:GNHAR-mat}
        A^{(\cdot)} = \operatorname{diag}(\alpha_{1}^{(\cdot)},\ldots,\alpha_{N}^{(\cdot)}) + \sum_{r=1}^{s^{(\cdot)}} \beta^{(\cdot)}_{r} W^{(r)}.
    \end{equation}
    We refer to this model as GNHAR($s^{(d)},s^{(w)},s^{(m)}$).
\end{definition}

Focusing on the daily component of \cref{eq:GNHAR-mat}, $\alpha_{i}^{(d)} \in \R$ represents the relevance of the $i$-th stock for $i=1,\ldots,N$, while $\beta^{(d)}_{r} \in \R$ encodes the effect of the $r$-stage neighbours for $r=1,\ldots, s^{(d)}$. The network order $s^{(d)} \in \Nat$ controls the maximal stage of neighbour dependence. Similar interpretations hold for $\alpha_{i}^{(w)},\beta^{(w)}_{r}, s^{(w)}$ and $\alpha_{i}^{(m)},\beta^{(m)}_{r}, s^{(m)}$ at the weekly and monthly aggregation levels, respectively.
As in the classical GNAR framework, we refer to the model as an individual-$\alpha$ GNHAR if the AR parameters are node-specific, or as a global-$\alpha$ GNHAR model if there is a common $\alpha^{(\cdot)}$ parameter shared among the components at each horizon. 

The GNHAR model naturally inherits the nearest-neighbour contribution to the network parameters and, consequently, handles missing data better than classical vector HAR models. Moreover, the AR component of \cref{eq:GNHAR} is characterised by $3N+s^{(d)}+s^{(w)}+s^{(m)}$ parameters in the individual-$\alpha$ case, or $3+s^{(d)}+s^{(w)}+s^{(m)}$ parameters in the global-$\alpha$ case, as opposed to the $3N^{2}$ parameters required by an unconstrained vector HAR model.

Our network-based HAR model is similar in spirit to recent contributions in the literature, such as \citet{zhang2025graph, tapia2025higher} for the HAR-DRD framework for realized covariances. However, the focus of our work is on forecasting a vector of RV processes. \cref{eq:GNHAR} is formulated using overlapping daily, weekly, and monthly averages as in the univariate HAR specification, and moreover it can accommodate directed graphs.

\subsection{The Jump-Continuous GNHAR Specification}
Our GNHAR framework can also be used to account for the jump and continuous components of the vector RV process separately, akin to \citet{andersen2007roughing} for the standard vector HAR model. To this end, we propose the \textit{JC-GNHAR model}.
Let $J_{i,t}$ be the jump component of the $i$-th asset defined according to \cref{eq:jump} for all $i=1,\ldots,N$ and $t\in\Z$, and let us denote by $LJ_{i,t} = \log(J_{i,t} + 1)$ the corresponding log-transformed quantity. Thus $\set{\mathbf{LJ}_{t}}_{t\in\Z}$ is the $N$-dimensional vector processes of the log-transform jump component, with $\mathbf{LJ}_{t} =(LJ_{1,t}, \ldots, JL_{N,t})^{\top}$ for all $t\in\Z$.
Likewise, let $C_{i,t}$ and $LC_{i,t} = \log(C_{i,t})$ indicate the continuous and the log-transformed continuous part of $RV_{i,t}$, for each $i=1,\ldots,N$ and $t\in\Z$, respectively; similarly $\set{\mathbf{LC}_{t}}_{t\in\Z}$ denotes the $N$-dimensional vector processes of the log-transform continuous component, where $\mathbf{LC}_{t} =(LC_{1,t}, \ldots, JC_{N,t})^{\top}$ for all $t\in\Z$. 

\begin{definition}[Generalised Network Heterogeneous Autoregressive Model with Jump-Continuous Decomposition]
    Given a graph $\G=(\V,\E)$, the GNHAR model of order ($s^{(d)},s^{(w)},s^{(m)}$) with the jump-continuos decomposition over $\G$ is
    \begin{equation}\label{eq:GNHAR-JC}
    \begin{aligned}
        \XX_{t} = & \; 
        \boldsymbol{\mu}+ A_{J}^{(d)} \LJ^{(d)}_{t-1} + A_{J}^{(w)} \LJ^{(w)}_{t-1} + A_{J}^{(m)} \LJ^{(m)}_{t-1} 
        + A_{C}^{(d)} \LC^{(d)}_{t-1} + A_{C}^{(w)} \LC^{(w)}_{t-1}  + A_{C}^{(m)} \LC^{(m)}_{t-1} + \vveps_{t}, 
    \end{aligned}
    \end{equation}
    for all $t \in \Z$. Here, $\mmu$ is the vector of intercept parameters, $\set{\vveps_t}_{t\in\Z}$ is a zero-mean Gaussian process with covariance matrix $\sigma^{2} I_{N}$, and
    $\LJ^{(\cdot)}_{t}$ and $\LC^{(\cdot)}_{t}$ are the jump and continuous aggregated components, respectively. The matrices $A_{J}^{(\cdot)}$ and $A_{C}^{(\cdot)}$ are defined as in \cref{eq:GNHAR-mat}, with $\alpha^{(\cdot)}_{J}$ and $\alpha^{(\cdot)}_{C}$ denoting the usual JC-HAR parameters from \cref{eq:JC-HAR}, and $\beta^{(\cdot)}_{J}$ and $\beta^{(\cdot)}_{C}$ denoting the network parameters for the respective jump and continuous component. We refer to this model as JC-GNHAR($s^{(d)},s^{(w)},s^{(m)}$).
\end{definition}

By distinguishing the continuous and jump parts of RV, we effectively double the number of predictors. Nonetheless, the parameter efficiency of our network-based approach allows to manage this expansion without producing noisy estimates when fitting the model. This enables a direct investigation of the interactions between the two components and among the assets by examining the estimated parameters. It facilitates the analysis of volatility spillovers, and disentangles whether these spillovers are primarily driven by continuous information flow or by sudden price jumps. Furthermore, by explicitly modelling the cross-connections through the graph, \cref{eq:GNHAR-JC} fully exploits the predictive power of both jumps and continuous volatility across different assets. This leads to more robust forecasts, particularly during periods of financial crisis. To the best of our knowledge, this is the first time a network-based approach has been proposed in this context.

\subsection{Financial Network Structures}
As both the proposed GNHAR and JC-GNHAR models require an a priori graph $\G=(\V,\E)$, we briefly review some of the most widely used networks in our financial context.

\paragraph{Fully Connected Graph.}
As a first example, we consider the fully connected graph. In this case, all assets are connected to all other assets, reflecting a uniform and static interdependence among the volatilities. Connections are assumed to be equally important and, despite its simplicity, this structure shows competitive results when forecasting RV data \citep{boetti2025long}.

\paragraph{Granger Causal Graph.}
Granger causality \citep{granger1969investigating} is commonly used to study cross-market spillover effects, as it helps determine whether a shock in one asset propagates to others. Formally, the process $\set{X_{i,t}}_{t\in\Z}$ is said to Granger-cause $\set{X_{j,t}}_{t\in\Z}$ if past values of the former contain information that improves forecasts of the latter. In such a case, the edge $i \rightsquigarrow j$ indicates a causal relation from the $i$-th component towards the $j$-th component. 

Once Granger causality tests have been performed for all pairs of stock indices, the directed edges of $\G$ are determined by the significant Granger-causal relationships. To account for multiple testing, we apply the Benjamini--Hochberg correction. Alternatively, to isolate spillover effects supported by stronger statistical evidence, the more conservative Bonferroni correction can be utilised to produce a sparser network.

\paragraph{Connectedness Measure Graph.}
The Diebold--Yilmaz connectedness index is a popular measure for quantifying the magnitude of volatility spillovers among assets \citep{diebold2009measuring, diebold2012better}. More precisely, let us consider two assets $i$ and $j$ and a forecast horizon $h > 0$. The (normalised) Diebold--Yilmaz connectedness index $d_{j,i}(h)$ represents the proportion of the $h$-step-ahead forecast error variance for asset $j$ that can be attributed to shocks in asset $i$. Hence, for each pair of components, we define the directional connectedness $i \rightsquigarrow j$ by evaluating $d_{j,i}(h)$ at a given $h$ \citep{diebold2014network}. 

In general, $d_{j,i}(h) \neq d_{i,j}(h)$, leading to the definition of a directed and weighted graph, whose edge weights are the connectedness indices. Note that, while $d_{j,i}(h) = 0$ indicates there is no spillover from asset $i$ towards asset $j$, it is rarely exactly zero in practice. Therefore, when selecting edges, we remove those whose corresponding connectedness indices are below a chosen threshold.

\section{Data Description and Experimental Design}\label{sec:data}
Our study focuses on the 5-min RV of $N=10$ stock market indexes: 2 from European Union, 4 from the United States of America, 1 from India, and 3 from East Asia; see \cref{tab:stock-ind} for more details. The data span August 6, 2013 to January 3, 2022, for a total of $T=2202$ trading days. Data were obtained from the Oxford-Man Institute of Quantitative Finance (\url{https://oxford-man.ox.ac.uk/research/realized-library}).
Descriptive statistics for the log-transformed RV time series are reported in \cref{tab:summary-stats-logRV} in Appendix \ref{app:additional}. Although the proposed methodology allows models to be fitted when missing values are present, we linearly interpolate any missing values as our aim is to compare the various models in terms of forecasting performance.

\begin{table}[htbp]
\centering
\begin{tabular}{l|l|l}
\toprule
\textbf{Ticker} & \textbf{Index} & \textbf{Country} \\
\midrule
DJI & Dow Jones Industrial Average & United States \\
GDAXI & DAX & Germany \\
HSI & Hang Seng Index & Hong Kong \\
IXIC & NASDAQ-100 & United States \\
KS11 & KOSPI (Korea Composite) & South Korea \\
N225 & Nikkei 225 & Japan \\
NSEI & NIFTY 50 & India \\
RUT & Russell 2000 & United States \\
SPX & S\&P 500 & United States \\
STOXX50E & EURO STOXX 50 & Eurozone \\
\bottomrule
\end{tabular}
\caption{Stock market index tickers, full names, and geographic locations.}
\label{tab:stock-ind}
\end{table}

We use a rolling window approach based on $S=1000$ days (approximately four trading years), and we consider both short- and long-horizon predictions, namely $h=1,5,10,22,44$, to assess the out-of-sample forecast accuracy of the models. This setup yields to a rolling estimation period August 6, 2013 to November 2, 2021, and an evaluation period from May 31, 2017 to January 3, 2022, with $1159$ estimations in total. 
In other words, for each rolling window $I_{t}=[t-S,t]$, where $t=S,\ldots,T-44$, we first identify the network $\G$ and fit the chosen model to the data. We then compute $h$-step-ahead forecasts for each stock market index, $\hat{\XX}_{t+h}$, with $h=1,5,10,22,44$. Note that by focusing on one rolling window $I_{t}$ at a time, we also incorporate time dynamics of the RV within the graphs.

We employ the direct approach to forecasts when $h>1$. This method consists in fitting a horizon-specific version of the model for each $h>1$, targeting $\XX_{t+h}$ directly. That is, for each horizon $h$ and each forecast origin $t$, we re-estimate the model on the window $I_{t}$ using~$\XX_{t-1+h}$ as the dependent variable, obtaining horizon-specific coefficient estimates. These estimates are then used to compute the direct $h$-step-ahead forecast $\hat{\XX}_{t+h}$ using information available up to time $t$. In contrast to the iterated approach, direct forecasts do not require the bi-power variation to be known, or estimated with a separate model, over the whole forecast horizon. Hence, we can equally compare the GNHAR predictions with those of JC-GNHAR. Moreover, direct forecasts are more robust to model misspecification \citep{marcellino2006comparison}.

\paragraph{Forecast Error Measure.}
Given the forecast error for each index $i=1,\ldots,N$ and at each horizon $h>0$
\begin{equation*}
    e_{i,t}(h) = X_{i,t+h} - \hat{X}_{i,t+h},
\end{equation*}
the overall performance of a chosen model is evaluated in terms of the mean absolute forecast error (MAFE), which is defined for each $t=S,\ldots,T-44$ and $h>0$ as 
\begin{equation*}
    \operatorname{MAFE}_{t}(h) = \frac{1}{N} \sum_{i=1}^{N} |e_{i,t}(h)|.
\end{equation*}
For each horizon $h>0$, the prediction performance of a model is then summarised by averaging the MAFEs over the considered times $t$, i.e.
\begin{equation*}
    \operatorname{avg-MAFE}(h) = \frac{1}{T-44-S+1} \sum_{t=S}^{T-44} \operatorname{MAFE}_{t}(h).
\end{equation*}
Although the mean squared forecast error is an alternative loss function that is consistent for the predictive mean \citep{gneiting2011making}, the MAFE is frequently preferred as it is more robust to extreme errors \citep{wilms2021multivariate, son2023forecasting}. The evaluation period considered in our study encompasses both tranquil intervals and periods of financial instability, such as the COVID-19 outbreak. Consequently, the average MAFE allows for a overall quantification of predictive accuracy, even in the presence of sudden structural changes.

\paragraph{Empirical Comparison of Financial Networks.}
Apart from the fully connected graph, the graphs based on Granger causality or the connectedness index naturally change throughout the estimation period.
To better investigate the cross-relationships among the assets, we define two types of Granger-causal graphs and two types of connectedness indices graphs. More precisely, for each rolling window, we find the graph based on the Granger causal relations with lag $p=1$ and $p=22$, at $5\%$ significance level and Benjamini--Hochberg correction. 
Similarly, we construct networks based on connectedness indices at predictive horizons $h=1$ and $h=22$, where we threshold edges with values below $0.05$. This means we discard connections where less than $5\%$ of the variability of one asset is explained by the variability of another asset. In the following, the notation ``GC1'' and ``GC22'' refer to the Granger causal graphs, and ``CI1'' and ``CI22'' to the connectedness index graphs.

\begin{figure}[H]
  \centering
  \includegraphics[width=0.9\linewidth]{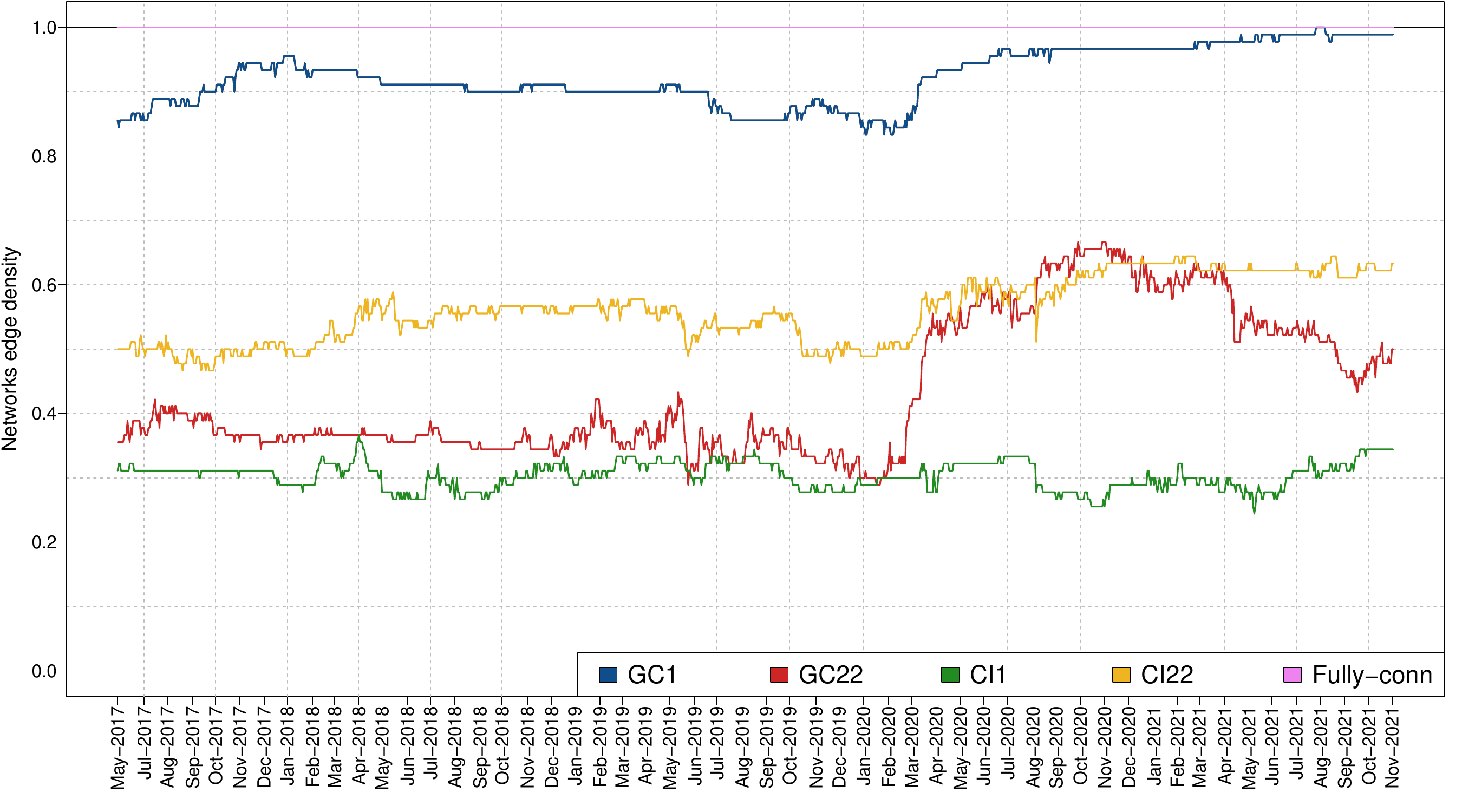}
  \caption{Edge density of the Granger causal graphs at either $p=1$ (blue) or $p=22$ (red), of the connectedness index graphs at either $h=1$ (green) or $h=22$ (yellow), and of the fully connected graph (violet).}
  \label{fig:edge-densities}
\end{figure}

\cref{fig:edge-densities} reports the edge densities of the networks, with reported dates corresponding to the final dates of each rolling window.
As expected, GC1 is denser than GC22, reflecting that Granger causal links are generally stronger over the immediate past ($p=1$) than over a longer historical period ($p=22$). A parallel intuition applies to the networks based on the connectedness index: CI22 exhibits larger density than CI1 because, at a longer predictive horizon ($h=22$), cross-asset spillovers have had more time to propagate throughout the system than in the immediate future ($h=1$).
Interestingly, GC22 remains relatively sparse before data from March 2020 are included, after which the graph becomes markedly denser. This pattern is consistent with the financial literature: spillover effects tend to intensify during crises, such as the COVID-19 outbreak, whereas in tranquil periods variation is more idiosyncratic \citep{diebold2012better}.
A similar, but less evident, behaviour can also be observed for GC1. Finally, the edge densities of CI1 and CI22 tend to remain constant, even though the edge weights change over time.

\clearpage
\section{Forecast Accuracy}\label{sec:forecast}

\subsection{Single-horizon Forecasts Results}
We start by investigating the forecasting performance of the network-based models for short-term prediction, specifically for $h=1$. For each chosen graph, we compare both model representations, namely GNHAR and JC-GNHAR approaches, using either global- or individual-$\alpha$ specifications and various combinations of network orders $(s^{(d)},s^{(w)},s^{(m)})$. Since, by definition, the fully connected graph has empty neighbourhood sets for any $s^{(\cdot)} > 1$, we restrict the network orders to 0 or 1 to ensure a fair comparison across the graphs. 
We also compare our network-based models with benchmarks, where no network structure is considered. These correspond to univariate HAR models, with or without the jump-continuous representation, fitted separately to each stock. \cref{tab:GNHAR-JCGNHAR-mafe-h1} reports the average MAFEs at horizon $h=1$ for the 28 model combinations across each of the five networks and the two benchmarks.

To validate differences in forecasting performance across the various configurations and to help discern the optimal model representation, we employ the Model Confidence Set (MCS) procedure \citep{hansen2011model}. Specifically, given a network, we compare the global-$\alpha$ GNHAR, individual-$\alpha$ GNHAR, global-$\alpha$ JC-GNHAR, and individual-$\alpha$ JC-GNHAR models across various network orders, alongside the HAR and JC-HAR benchmarks. The MCS is evaluated in \texttt{R} using the \texttt{MCSprocedure} function \citep{MCS}. We use the TR test statistic based on 1,000 bootstrap samples at a $20\%$ significance level. This analysis is conducted for each of the five graphs, but the MCS results for the HAR and JC-HAR benchmarks are reported only once in \cref{tab:GNHAR-JCGNHAR-mafe-h1} because they did not change across the five runs.

Notably, the global-$\alpha$ GNHAR and JC-GNHAR models outperform the benchmarks across all networks and network orders. In particular, the global-$\alpha$ JC-GNHAR configuration significantly improves forecasting performance compared to the JC-HAR benchmark, achieving up to a $38\%$ reduction in the average MAFE. 
The same does not hold for the individual-$\alpha$ specification. Indeed, the individual-$\alpha$ models appear less stable and more sensitive to the choice of network and network order, and occasionally yield worse forecasts than the benchmarks.
Overall, both the benchmarks and individual-$\alpha$ models are consistently excluded from the MCS. This highlights the importance of capturing dependencies between components via the network. Moreover, the parameter parsimony offered by the global-$\alpha$ specification is preferable for accurate forecasting, and it is particularly relevant when employing the jump-continuous decomposition.

The CI22-based models appear to be the best-performing models overall and, in fact, the two lowest average MAFEs are achieved by the global-$\alpha$ JC-GNHAR on such a network. Remarkably, the GNHAR modelling approach is best for the fully connected, GC1, and GC22 graphs, as opposed to the jump-continuous representation which appears more suitable for the weighted graphs CI1 and CI22.

While the most suitable network order for predicting the RV depends naturally on the network used, we can still observe some interesting patterns across the networks. More precisely, the orders (1,0,0), (0,1,0), (0,0,1), (0,1,1), and (1,1,1) lead to higher MAFE losses, whereas models with network order (1,0,1) yield the lowest (or tied for the lowest) MAFE values under both the global- and individual-$\alpha$ settings.
Models with a (1,1,0) order perform very similarly under the global-$\alpha$ specification, though they are less stable when paired with individual-$\alpha$ parameters, especially under the JC-GNHAR representation. Altogether, the daily network component appears to be crucial, and a single network component alone may not be enough for one-step-ahead predictions. On the other hand, it is better to incorporate only one additional network component rather than including both the weekly and monthly network components. This is further confirmed by the MCS procedure, as (1,1,0) and (1,0,1) are the only orders always included in the set.

Finally, we replicate the above analysis using the mean squared forecast error (MSFE) as the loss function. The results, detailed in \cref{tab:GNHAR-JCGNHAR-msfe-h1} of Appendix \ref{app:additional}, confirm our primary findings and show that that the MCS procedure fundamentally identifies the same models.

\begin{table}[htbp]
\centering
{\setlength{\tabcolsep}{9pt}\small
\begin{tabular}{ll|cccc}
\toprule
& & \multicolumn{2}{c}{\textbf{GNHAR}} & \multicolumn{2}{c}{\textbf{JC-GNHAR}} \\
\textbf{Network} & \textbf{Order} & \textbf{gl-}$\boldsymbol{\alpha}$ & \textbf{ind-}$\boldsymbol{\alpha}$ & \textbf{gl-}$\boldsymbol{\alpha}$ & \textbf{ind-}$\boldsymbol{\alpha}$ \\
\midrule
\multirow{7}{*}{Fully} & (1,0,0) & 0.440 (<0.01) & 0.602 (<0.01) & 0.446 (<0.01) & 0.760 (<0.01) \\
& (0,1,0) & 0.440 (<0.01) & 0.606 (<0.01) & 0.438 (0.04) & 0.836 (<0.01) \\
& (0,0,1) & 0.456 (<0.01) & 0.609 (<0.01) & 0.469 (<0.01) & 0.798 (<0.01) \\
& (0,1,1) & 0.439 (<0.01) & 0.508 (<0.01) & 0.452 (0.01) & 0.688 (<0.01) \\
& (1,0,1) & \textbf{0.433 (1)} & 0.479 (<0.01) & 0.451 (<0.01) & 0.547 (<0.01) \\
& (1,1,0) & \textbf{0.433 (0.36)} & 0.540 (<0.01) & 0.440 (0.04) & 0.805 (<0.01) \\
& (1,1,1) & 0.452 (0.05) & 0.503 (<0.01) & 0.452 (0.02) & 0.643 (<0.01) \\
\midrule
\multirow{7}{*}{GC1} & (1,0,0) & 0.440 (<0.01) & 0.599 (<0.01) & 0.447 (<0.01) & 0.754 (<0.01) \\
& (0,1,0) & 0.440 (<0.01) & 0.597 (<0.01) & 0.438 (0.02) & 0.823 (<0.01) \\
& (0,0,1) & 0.459 (<0.01) & 0.607 (<0.01) & 0.468 (<0.01) & 0.798 (<0.01) \\
& (0,1,1) & 0.439 (<0.01) & 0.505 (<0.01) & 0.453 (<0.01) & 0.692 (<0.01) \\
& (1,0,1) & \textbf{0.433 (1)} & 0.478 (<0.01) & 0.453 (<0.01) & 0.547 (<0.01) \\
& (1,1,0) & \textbf{0.434 (0.34)} & 0.537 (<0.01) & 0.440 (0.02) & 0.806 (<0.01) \\
& (1,1,1) & 0.459 (0.02) & 0.503 (<0.01) & 0.450 (0.01) & 0.644 (<0.01) \\
\midrule
\multirow{7}{*}{GC22} & (1,0,0) & 0.440 (<0.01) & 0.576 (<0.01) & 0.441 (0.04) & 0.740 (<0.01) \\
& (0,1,0) & 0.440 (<0.01) & 0.546 (<0.01) & 0.441 (0.04) & 0.773 (<0.01) \\
& (0,0,1) & 0.473 (<0.01) & 0.583 (<0.01) & 0.460 (<0.01) & 0.770 (<0.01) \\
& (0,1,1) & 0.459 (<0.01) & 0.489 (<0.01) & 0.454 (<0.01) & 0.643 (<0.01) \\
& (1,0,1) & \textbf{0.435 (1)} & 0.473 (<0.01) & 0.443 (0.05) & 0.598 (<0.01) \\
& (1,1,0) & \textbf{0.436 (0.23)} & 0.537 (<0.01) & 0.441 (0.08) & 0.757 (<0.01) \\
& (1,1,1) & 0.486 (<0.01) & 0.531 (<0.01) & 0.461 (<0.01) & 0.656 (<0.01) \\
\midrule
\multirow{7}{*}{CI1} & (1,0,0) & 0.439 (0.02) & 0.582 (<0.01) & 0.437 (0.13) & 0.693 (<0.01) \\
& (0,1,0) & 0.440 (0.01) & 0.573 (<0.01) & 0.439 (0.02) & 0.779 (<0.01) \\
& (0,0,1) & 0.467 (<0.01) & 0.632 (<0.01) & 0.462 (<0.01) & 0.876 (<0.01) \\
& (0,1,1) & 0.443 (0.09) & 0.546 (<0.01) & 0.443 (0.13) & 0.689 (<0.01) \\
& (1,0,1) & 0.437 (0.13) & 0.466 (<0.01) & \textbf{0.435 (1)} & 0.579 (<0.01) \\
& (1,1,0) & 0.438 (0.09) & 0.561 (<0.01) & \textbf{0.436 (0.27)} & 0.739 (<0.01) \\
& (1,1,1) & 0.454 (<0.01) & 0.564 (<0.01) & 0.459 (0.01) & 0.617 (<0.01) \\
\midrule
\multirow{7}{*}{CI22} & (1,0,0) & 0.436 (<0.01) & 0.601 (<0.01) & 0.435 (<0.01) & 0.723 (<0.01) \\
& (0,1,0) & 0.440 (<0.01) & 0.582 (<0.01) & 0.437 (<0.01) & 0.805 (<0.01) \\
& (0,0,1) & 0.466 (<0.01) & 0.602 (<0.01) & 0.480 (<0.01) & 0.844 (<0.01) \\
& (0,1,1) & 0.439 (<0.01) & 0.524 (<0.01) & 0.436 (<0.01) & 0.675 (<0.01) \\
& (1,0,1) & \textbf{0.433 (0.45)} & 0.452 (<0.01) & \textbf{0.432 (0.97)} & 0.519 (<0.01) \\
& (1,1,0) & \textbf{0.434 (0.27)} & 0.544 (<0.01) & \textbf{0.432 (1)} & 0.766 (<0.01) \\
& (1,1,1) & \textbf{0.439 (0.36)} & 0.482 (<0.01) & \textbf{0.441 (0.24)} & 0.570 (<0.01) \\
\midrule
\multicolumn{2}{c|}{Benchmark} & \multicolumn{2}{c}{0.504 (<0.01)} & \multicolumn{2}{c}{0.692 (<0.01)} \\
\bottomrule
\end{tabular}}
\caption{Average MAFEs at horizon $h=1$ for GNHAR and JC-GNHAR models. MCS test $p$-values are in brackets. Bold indicates inclusion in the MCS at $20\%$ significance.} \label{tab:GNHAR-JCGNHAR-mafe-h1}
\end{table}

\subsection{Multi-horizon Forecasts Results}
Following the results on the short-term predictions, we investigate the forecasting performance at longer horizons only for those models consistently not discarded by the MCS tests. Specifically, we focus on global-$\alpha$ GNHAR and JC-GNHAR models with network orders (1,0,1) or (1,1,0) and we compare them against the HAR and JC-HAR benchmarks. The average MAFEs at $h=5, 10, 22, 44$ are presented in \cref{tab:GNHAR-JCGNHAR-mafe-hlong}.

Once again, network models demonstrate superior predictive performance over both benchmarks, with the performance gap widening as the forecast horizon increases. 
Indeed, at $h=5$, the best network model outperforms the best benchmark by $26\%$, and at $h=44$ there is a $40\%$ improvement in the average MAFE.
Moreover, as the forecasting horizon extends, we see the predictive power of the non-network models diminishes rapidly, while the network models exhibit greater stability.

The performance of the fully connected graph degrades as the horizon expands, and a similar behaviour can be observed for the networks based on one-period dependencies. Indeed, the network-based models using GC1 and CI1 are competitive at $h=5$, but they are outperformed by their GC22 and CI22 counterparts at horizons $h=22$ and $h=44$. This reflects the fact that the 22-period graphs are more effective at filtering out daily noise, thereby capturing more persistent systemic relationships. Furthermore, these networks exhibit intermediate sparsity compared to the other three, potentially making the associated GNHAR and JC-GNHAR models more robust for long-term forecasting.

We further observe a distinct time-horizon pattern between the (1,1,0) and (1,0,1) specifications under the GNHAR framework. At $h=5$ and $h=10$, the (1,1,0) order frequently dominates the (1,0,1) order across networks, whereas the opposite holds at $h=22$ and $h=44$. When employing the JC decomposition, the (1,1,0) order generally leading to better predictions than (1,0,1) order only at $h=5$. This suggests that the monthly network component is effective at capturing the long-term dynamics, which may become particularly important when forecasting over extended horizons.

These conclusions remain generally valid when using the MSFE (see \cref{tab:GNHAR-JCGNHAR-msfe-hlong} in Appendix~\ref{app:additional}). Interestingly, we observe that the GNHAR(1,0,1) model on the fully connected graph achieves the lowest average MSFE at $h=44$. However, since this performance is not attained at other horizons, it may be attributable to the sensitivity of the MSFE loss function to extreme errors.

\begin{table}[!ht]
\centering
{\setlength{\tabcolsep}{9pt}\small
\begin{tabular}{ll|cccc}
\toprule
\textbf{Network} & \textbf{Model} & $\boldsymbol{h=5}$ & $\boldsymbol{h=10}$ & $\boldsymbol{h=22}$ & $\boldsymbol{h=44}$ \\
\midrule
\multirow{4}{*}{Fully} & GNHAR(1,0,1) & 0.570 & 0.640 & 0.745 & 0.796 \\
& GNHAR(1,1,0) & 0.561 & 0.657 & 0.767 & 0.801 \\
& JC-GNHAR(1,0,1) & 0.593 & 0.667 & 0.774 & 0.769 \\
& JC-GNHAR(1,1,0) & 0.552 & 0.671 & 0.779 & 0.865 \\
\midrule
\multirow{4}{*}{GC1} & GNHAR(1,0,1) & 0.569 & 0.633 & 0.730 & 0.820 \\
& GNHAR(1,1,0) & 0.565 & 0.652 & 0.726 & 0.787 \\
& JC-GNHAR(1,0,1) & 0.602 & 0.654 & 0.761 & 0.773 \\
& JC-GNHAR(1,1,0) & 0.556 & 0.665 & 0.788 & 0.857 \\
\midrule
\multirow{4}{*}{GC22} & GNHAR(1,0,1) & 0.578 & 0.659 & 0.732 & 0.776 \\
& GNHAR(1,1,0) & 0.567 & 0.644 & 0.802 & 0.841 \\
& JC-GNHAR(1,0,1) & 0.599 & 0.637 & 0.707 & 0.744 \\
& JC-GNHAR(1,1,0) & 0.566 & 0.673 & 0.812 & 0.829 \\
\midrule
\multirow{4}{*}{CI1} & GNHAR(1,0,1) & 0.565 & 0.695 & 0.685 & 0.753 \\
& GNHAR(1,1,0) & 0.555 & 0.637 & 0.723 & 0.813 \\
& JC-GNHAR(1,0,1) & 0.559 & 0.637 & 0.696 & 0.748 \\
& JC-GNHAR(1,1,0) & \textbf{0.550} & 0.655 & 0.737 & 0.773 \\
\midrule
\multirow{4}{*}{CI22} & GNHAR(1,0,1) & 0.558 & 0.628 & \textbf{0.678} & 0.747 \\
& GNHAR(1,1,0) & 0.557 & 0.631 & 0.691 & 0.787 \\
& JC-GNHAR(1,0,1) & 0.554 & \textbf{0.620} & 0.707 & \textbf{0.739} \\
& JC-GNHAR(1,1,0) & \textbf{0.550} & 0.622 & 0.683 & 0.770 \\
\midrule
\multicolumn{2}{c|}{HAR Benchmark} & 0.692 & 0.798 & 1.200 & 1.237 \\
\multicolumn{2}{c|}{JC-HAR Benchmark} & 0.746 & 0.883 & 1.280 & 1.302 \\
\bottomrule
\end{tabular}}
\caption{Average MAFEs at horizons $h=5, 10, 22, 44$ for the selected GNHAR and JC-GNHAR models. Bold font highlights the best predictive performance at each horizon.} \label{tab:GNHAR-JCGNHAR-mafe-hlong}
\end{table}

\subsection{Temporal Forecast Error Dynamics}
By comparing the performance across all horizons, the JC-GNHAR(1,0,1) model with the CI22 graph yields the most accurate predictions overall. The GNHAR(1,0,1) model with the CI22 network is also a good alternative, especially for long-term forecasts. This suggests not only that exploiting a financial structure is better than a naive, fully connected representation, but also that weighted network can better represent connections between assets compared to binary indicators of Granger-causal relationships.

We illustrate the performance over time of the global-$\alpha$ JC-GNHAR(1,0,1) model over the CI22 network. For each horizons $h=1, 5, 10, 22, 44$, the relative absolute forecast error averaged across all stocks is presented in \cref{fig:JCGNHAR101-MAFE-time} in Appendix~\ref{app:additional}.

As expected, the largest absolute forecast errors at horizon $h=1$, \cref{fig:JCGNHAR101-MAFE-time}(a), coincide with major macroeconomic shocks. For example, these occur during the first and second waves of the COVID-19 pandemic, namely March 2020 and the period from November 2020 to January 2021. The highest error was registered on March 12, 2020, widely known as the `Black Thursday' of the pandemic crash. The two major market instabilities periods in 2018 are also captured, as we can see from distinct spikes during the February `Volmageddon' event and during the fourth quarter. Another relevant date is January 27, 2021, which reflects the turbulence caused by the anomalous GameStop short squeeze.

When examining the long-term absolute forecast errors in \cref{fig:JCGNHAR101-MAFE-time}(b)--(e), the periods of financial turmoil become even more pronounced. The first wave of the COVID-19 pandemic dominates the time-varying dynamics, particularly at the extended horizons of $h=22$ and $h=44$. Additionally, 2018 clearly emerges as another period of high financial instability, as the magnitude of the absolute forecast errors is notably larger than the overall average.

\subsection{Jump-Continuous Decomposition and Network Choice}
We now examine \cref{tab:GNHAR-JCGNHAR-mafe-h1} and \cref{tab:GNHAR-JCGNHAR-mafe-hlong} in more detail to better understand the role of the graph structure when combined with the jump-continuous decomposition.
We use the Diebold-Mariano (DM) test \citep{diebold1995comparing} to assess whether the two different forecasting models have significantly different levels of predictive accuracy. In particular, we utilise the modified version proposed by \citet{harvey1997testing} with variance estimator based on Bartlett weights, as implemented in \texttt{R} with \texttt{dm.test} function \citep{forecast}. 

In the absence of a network structure, the jump-continuous representation leads to a worse average performance than the standard HAR benchmark across all horizons. However, when paired with a graphical structure, models with the jump-continuous decomposition yield the best performance. For instance, the global-$\alpha$ JC-GNHAR framework over the CI22 graph achieves the smallest average MAFE for $h=1,5,10,44$. Thus, the network topology enables the model to better capture how jumps propagate through the system, converting immediate shocks into useful predictive information.

Comparing the results of the GNHAR and JC-GNHAR models reveals a trade-off between accuracy and robustness.
Under misspecified configurations, such as the individual-$\alpha$ specification, the average MAFEs at $h=1$ (see \cref{tab:GNHAR-JCGNHAR-mafe-h1}) show GNHAR models are typically stable, as opposed to JC-based models. This is further supported by the DM tests, which consistently confirm the statistical superiority of the GNHAR over the JC-GNHAR at the $5\%$ significance level for every given network and order.
Conversely, when the modelling configuration is more suitable, such as the global-$\alpha$ specification, the statistical superiority of the GNHAR over the JC-GNHAR framework is not always supported by the DM test. In fact, under the GC22 network, DM tests often suggest no significant difference between the two approaches, whereas for the CI1 and CI22 graphs, the JC-GNHAR model occasionally outperforms its GNHAR counterpart (e.g., CI1 graph with (1,0,1) or (1,1,0) orders, $p\text{-values}<0.01$; or CI22 graph with a (1,1,0) order, $p\text{-value}=0.03$).

At $h=5$, the JC-GNHAR(1,1,0) models over either the CI1 or CI22 networks outperform their respective GNHAR(1,1,0) counterparts ($p\text{-value}= 0.04$ and $p\text{-value}=0.03$, respectively). However, as the forecast horizon increases, DM tests do not support a significant difference between the two approaches. Therefore, explicitly accounting for the continuous and jump components enables the extraction of more useful information from this dataset, provided that the underlying network structure is appropriate and the model remains parsimonious. In such cases, we can obtain better predictions, particularly at short-term horizons. Otherwise, the standard GNHAR framework is preferable as it is generally more stable.

\section{Including Option-Implied Variance in GNHAR Models}\label{sec:exogenous}
The GNHAR model \eqref{eq:GNHAR} successfully captures both persistence and the cross-component relationships of asset RVs, but it is not designed to incorporate external market information, such as option-implied variance (IV). As a market-based measure of expected future return variability extracted from option prices, the role of IV in forecasting RV has been extensively studied and it is generally found to improve predictive performance \citep{christensen1998relation, britten2000option}.
In contrast to RV, which is based on high-frequency intraday returns, IV summarises market-implied (risk-neutral) expectations about volatility over a future horizon and may therefore provide information beyond lagged RV. Consequently, IV can be viewed as a forward-looking indicator of upcoming stock market volatility.
See, among others, \citet{blair2001forecasting, poon2003forecasting}, and \citet{bekaert2014vix} for a review. Applications of option-implied volatility as an exogenous predictor in RV forecasting models, including multivariate settings, can be found in \citet{busch2011role, liang2020implied, wilms2021multivariate}.

We define the GNHARX model below in a similar spirit to \cite{nason2022quantifying} for the GNAR model, which allows us to effectively embed the IV process within a network-based modelling framework.

\begin{definition}[Generalized Network Heterogeneous Autoregressive Model with Exogenous Variable]
    Given a graph $\G=(\V,\E)$, let $\set{\YY_{t}}_{t\in\Z}$ be a $N$-dimensional process of node-specific exogenous time series. The GNHAR model of order ($s^{(d)},s^{(w)},s^{(m)}$) with exogenous variable $\YY_{t}$ of order $p$ over $\G$ is defined as
    \begin{equation}\label{eq:GNHARX}
        \XX_{t} = \boldsymbol{\mu}+ A^{(d)} \XX^{(d)}_{t-1} + A^{(w)} \XX^{(w)}_{t-1} + A^{(m)} \XX^{(m)}_{t-1} + \sum_{j=1}^{p} \Lambda_{j} \YY_{t-j} + \vveps_{t}
    \end{equation}
    for all $t\in\Z$. The exogenous parameter matrix is $\Lambda_{j} = \operatorname{diag}\left( \lambda_{1,j}, \ldots, \lambda_{N,j} \right)$, $\mmu$ is the vector of intercept parameters, and $\vveps=\set{\vveps_t}_{t\in\Z}$ is a zero-mean Gaussian process with covariance matrix $\sigma^{2} I_{N}$.
    We refer to this model as GNHARX([$s^{(d)},s^{(w)},s^{(m)}$],$p$).
\end{definition}

The parameter $\lambda_{i,j}$ measures the strength of the relationship between the $i$-th exogenous component and the process at lag $j$.
Since the exogenous parameters are node-specific, \cref{eq:GNHARX} describes the \emph{local}-$\lambda$ model and, similarly, we have the \emph{global}-$\lambda$ model when $\lambda_{i,j}=\lambda_{j}$ for all~$i=1,\ldots,N$ and $j=1,\ldots,p$. 

\paragraph{The Daily-Weekly-Monthly Exogenous Model.}
We further generalise the \cref{eq:GNHARX} to allow the exogenous processes to interact over the network and across time horizons. In the same spirit of particular, we investigate a daily-weekly-monthly (dwm) network representation of the exogenous component. Specifically, we define the GNHARX([$s^{(d)},s^{(w)},s^{(m)}$], [$z^{(d)},z^{(w)},z^{(m)}$]) model as
\begin{equation}\label{eq:GNHARX-dwm}
\begin{aligned}
    \XX_{t} = \boldsymbol{\mu}
    + A^{(d)} \XX^{(d)}_{t-1} + A^{(w)} \XX^{(w)}_{t-1} + A^{(m)} \XX^{(m)}_{t-1} 
    + \Lambda^{(d)} \YY^{(d)}_{t-1} + \Lambda^{(w)} \YY^{(w)}_{t-1} + \Lambda^{(m)} \YY^{(m)}_{t-1} + \vveps_{t},
\end{aligned}
\end{equation}
where $\YY^{(d)}_{t}$, $\YY^{(w)}_{t}$, and $\YY^{(m)}_{t}$ are the daily, weekly, and monthly aggregated exogenous components, respectively. The exogenous coefficient matrices are
\begin{equation}\label{eq:GNHARX-dwm-lambda}
    \Lambda^{(\cdot)} = \operatorname{diag}(\lambda_{1}^{(\cdot)},\ldots,\lambda_{N}^{(\cdot)}) + \sum_{r=1}^{z^{(\cdot)}} \nu^{(\cdot)}_{r} W^{(r)}.
\end{equation}
Hence, $\lambda_{i}^{(d)}$ controls the $i$-th direct daily exogenous effect, $\nu^{(d)}_{r}$ controls the daily exogenous effect transmitted through the $r$-stage neighbours of $\G$, and $z^{(d)}$ is the maximal network stage at the daily level. The same interpretation applies to the weekly and monthly components.

\begin{remark}
While \cref{eq:GNHARX} and \cref{eq:GNHARX-dwm} have been defined to account for only one exogenous regressor, they can naturally be extended to incorporate multiple exogenous processes when available. More precisely, let $\set{\YY_{k,t}}_{t\in\Z}$ be the $N$-dimensional process relative to the $k$-th exogenous, and let $p_{k}$ denote its AR order, for $k=1,\ldots,K\in\Nat$.
The GNHARX model([$s^{(d)},s^{(w)},s^{(m)}$],$\boldsymbol{p}$), with exogenous AR order $\boldsymbol{p}=(p_{1},\ldots,p_{K})$, over $\G$ is defined as
\begin{equation*}
    \XX_{t} = \boldsymbol{\mu}+ A^{(d)} \XX^{(d)}_{t-1} + A^{(w)} \XX^{(w)}_{t-1} + A^{(m)} \XX^{(m)}_{t-1} + \sum_{k=1}^{K} \sum_{j=1}^{p} \Lambda_{k,j} \YY_{k, t-j} + \vveps_{t},
\end{equation*}
for all $t\in\Z$. 
In particular, $\Lambda_{k,j} = \operatorname{diag}\left( \lambda_{1,k,j}, \ldots, \lambda_{N,k,j} \right)$ is the parameter matrix associated to the $k$-th exogenous regressor. \newline
Similarly, the dwm-representation of the GNHARX model([$s^{(d)},s^{(w)},s^{(m)}$], [$\boldsymbol{z}^{(d)},\boldsymbol{z}^{(w)},\boldsymbol{z}^{(m)}$]) is
\begin{equation*}
\begin{aligned}
    \XX_{t} = \boldsymbol{\mu}
    + A^{(d)} \XX^{(d)}_{t-1} + A^{(w)} \XX^{(w)}_{t-1} + A^{(m)} \XX^{(m)}_{t-1} 
    + \sum_{k=1}^{K} \left( \Lambda^{(d)}_{k} \YY^{(d)}_{k,t-1} + \Lambda^{(w)}_{k} \YY^{(w)}_{k,t-1} + \Lambda^{(m)}_{k} \YY^{(m)}_{k,t-1} \right) + \vveps_{t}.
\end{aligned}
\end{equation*}
For a given daily, weekly, or monthly dependence, $\Lambda^{(\cdot)}_{k}$ is the parameter matrix associated to the $k$-th exogenous regressor, each defined as \cref{eq:GNHARX-dwm-lambda}. Note the maximal network stages of the exogenous regressors are encoded with the vectors $\boldsymbol{z}^{(\cdot)} = (z^{(\cdot)}_{1},\ldots, z^{(\cdot)}_{K})$. 
\end{remark}

\subsection{Forecast Comparative Analysis}
We investigate the impact of the IV component has on the forecasting performance across various time horizons and different network structures and network orders. The option-implied variances used in our analysis have been downloaded from Investing.com (\url{https://www.investing.com/}) and FRED (\url{https://fred.stlouisfed.org/}), and descriptive statistics for the log-transformed IV time series are reported in \cref{tab:summary-stats-2logIV} in Appendix \ref{app:additional}. Missing values have been processed processed similarly to the RV data.

Building on the results in \cref{sec:forecast}, we adopt the global-$\alpha$ GNHAR(1,0,1) model over the CI22 network as our baseline, with or without the JC decomposition. We fit GNHARX([1,0,1],$\cdot$) models under the global-$\lambda$ specifications, and using an AR(1) or a dwm representation for the exogenous process. We report results only for the AR(1) specification, as we empirically observed that higher-order lags lead to numerical instability. 
Similarly, the individual-$\lambda$ specification yields larger errors than its global-$\lambda$ counterpart, and this discrepancy becomes more pronounced for more complex models, for instance when the dwm representation of the exogenous component is employed. Since, this behaviour aligns with our earlier findings regarding the $\alpha$ specifications, we do not present the MAFEs relative to the individual-$\lambda$ models, although they are available from the authors upon request.
According to our previous analysis, we only consider network orders $[z^{(d)}, z^{(w)}, z^{(m)}]$ including the daily component for the exogenous regressor. These models are compared against the non-network benchmark, which corresponds to fitting univariate HARX models to each stock individually. The average MAFE losses across various horizons are reported in \cref{tab:GNHARX-mafe}, and  \cref{tab:GNHARX-msfe} in Appendix \ref{app:additional} repoprts the average MSFEs.

The GNHARX with exogenous order AR(1), and with the dwm network orders [0,0,0] or [1,0,0], outperform the HARX(1) benchmark across all horizons (DM tests with $p\text{-values}<0.01$). In fact, the GNHARX([1,0,1],1) model improves upon the average performance of the HARX(1) benchmark by at least $30\%$ over the forecast horizons, reaching an improvement of $44\%$ for the longest one. 
On the other hand, modelling IV with extensive network interactions does not always yield predictive benefits. While the [1,0,1] and [1,1,0] specifications are generally significantly better than the benchmark at short and medium horizons, this is not mirrored for the long-term forecasts. When compared to the [1,1,1] specification, the benchmark provides superior forecasts at $h=1$ (DM test $p\text{-value}<0.01$). Analogous conclusions hold for the forecasting performance of the JC-based models.

Overall, the GNHARX([1,0,1],1) model appears to be the most effective approach for forecasting when IV is included as an exogenous regressor. Apart from $h=5$, where the DM test indicates that the GNHARX model outperforms the respective JC-GNHARX model ($p\text{-value}= 0.02$), there is no an optimal modelling frameworks between these two. Nevertheless, it is clear that incorporating the exogenous regressor within the network-based approach leads to better predictions primarily at short-term horizons, as indicated by the GNHARX model outperforming the GNHAR model at $h=1$ ($p\text{-value}= 0.08$). For longer prediction, instead, both the GNHAR and JC-GNHAR models are significantly superior at longer horizons. This is also confirmed by the DM tests with $p\text{-values}< 0.08$ at $h=10, 22,44$. Note this is reasonable given the nature of unconditional forecasting and the presence of abrupt shifts within the considered data, and similar behaviour was also observed by \citet{wilms2021multivariate}.

\begin{table}[H]
\centering
{\setlength{\tabcolsep}{7pt}\small
\begin{tabular}{lc|ccccc}
\toprule
\textbf{Model} & \textbf{ARX Order} & $\boldsymbol{h=1}$ & $\boldsymbol{h=5}$ & $\boldsymbol{h=10}$ & $\boldsymbol{h=22}$ & $\boldsymbol{h=44}$ \\
\midrule
HAR Benchmark & 0 & 0.504 & 0.692 & 0.798 & 1.200 & 1.237 \\
JC-HAR Benchmark & 0 & 0.692 & 0.746 & 0.883 & 1.280 & 1.302 \\
\midrule
GNHAR & 0 & 0.433 & 0.558 & 0.628 & \textbf{0.678} & 0.747 \\
JC-GNHAR & 0 & 0.432 & \textbf{0.554} & \textbf{0.620} & 0.707 & \textbf{0.739} \\
\midrule
\multirow{6}{*}{gl-$\lambda$ GNHARX} & 1 & 0.430 & 0.566 & 0.649 & 0.746 & 0.791 \\
& [0,0,0] & 0.434 & 0.697 & 0.812 & 0.827 & 0.924 \\
& [1,0,0] & 0.443 & 0.645 & 0.768 & 0.966 & 1.015 \\
& [1,0,1] & 0.578 & 0.726 & 1.002 & 1.168 & 1.174 \\
& [1,1,0] & 0.606 & 0.673 & 0.826 & 1.235 & 1.323 \\
& [1,1,1] & 0.729 & 0.849 & 1.309 & 1.411 & 1.255 \\
\midrule
\multirow{6}{*}{gl-$\lambda$ JC-GNHARX} & 1 & 0.430 & 0.581 & 0.663 & 0.754 & 0.789 \\
& [0,0,0] & \textbf{0.424} & 0.654 & 0.814 & 0.851 & 0.952 \\
& [1,0,0] & 0.440 & 0.661 & 0.790 & 0.862 & 1.079 \\
& [1,0,1] & 0.519 & 0.724 & 1.172 & 0.964 & 1.125 \\
& [1,1,0] & 0.628 & 0.668 & 0.810 & 1.159 & 1.394 \\
& [1,1,1] & 0.698 & 0.860 & 1.001 & 1.129 & 1.266 \\
\midrule
HARX Benchmark & 1 & 0.614 & 0.918 & 1.041 & 1.234 & 1.407 \\
JC-HARX Benchmark & 1 & 0.614 & 0.984 & 1.172 & 1.323 & 1.488 \\
\bottomrule
\end{tabular}}
\caption{Average MAFEs at horizons $h=1, 5, 10, 22, 44$ for the GNHARX models. Bold font highlights the best predictive performance at each horizon.} \label{tab:GNHARX-mafe}
\end{table}

\subsection{Estimated Model Coefficients}
\begin{figure}[htbp]
\centering
\begin{subfigure}{0.7\textwidth}
\includegraphics[width=\linewidth]{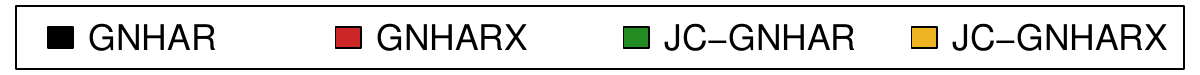}
\end{subfigure}
\vspace{0.5cm}

\begin{subfigure}{0.49\textwidth}
\includegraphics[width=\linewidth]{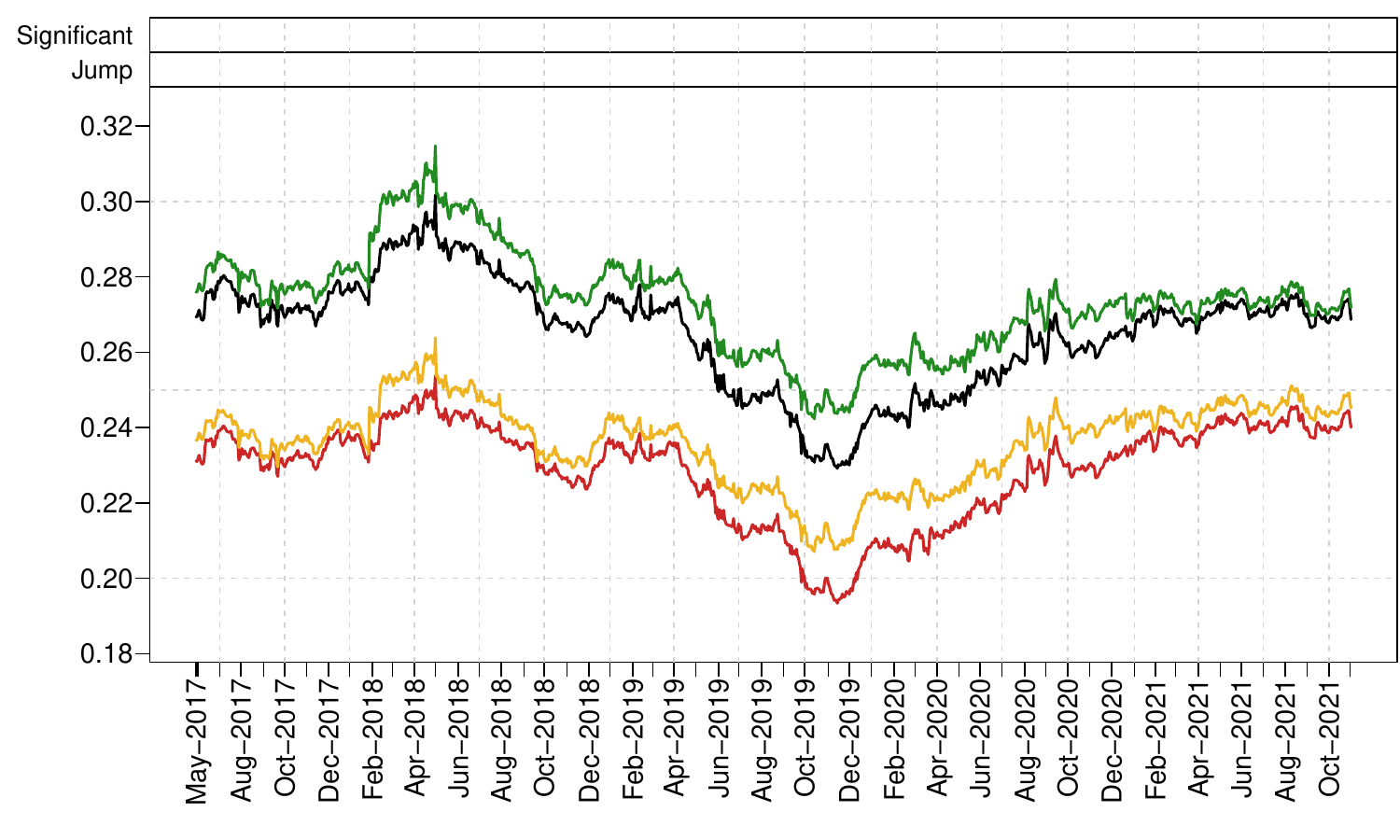}
\caption{$\hat\alpha^{(d)}$}
\end{subfigure}
\hfill
\begin{subfigure}{0.49\textwidth}
\includegraphics[width=\linewidth]{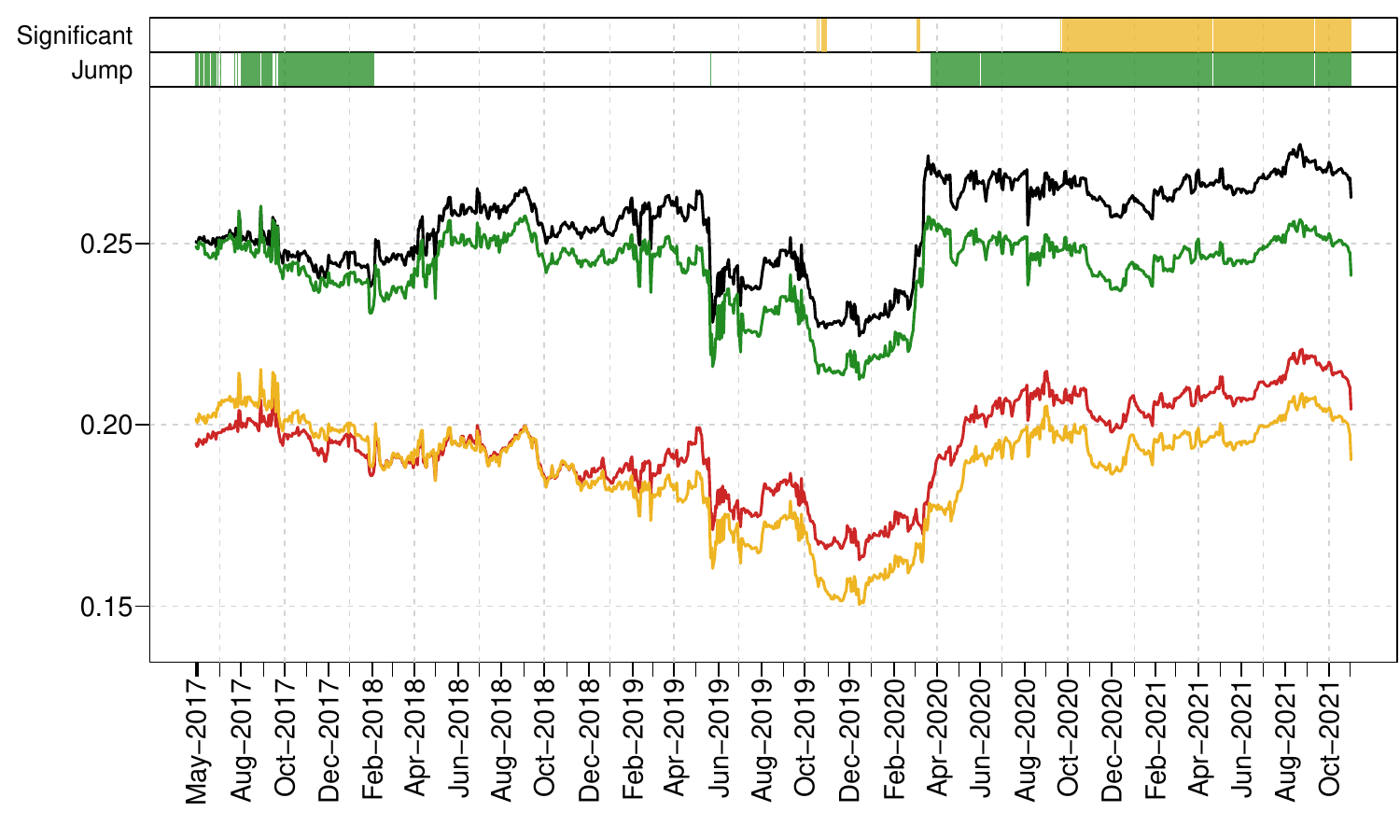}
\caption{$\hat\beta_{1}^{(d)}$}
\end{subfigure}
\vspace{0.5cm}

\begin{subfigure}{0.49\textwidth}
\includegraphics[width=\linewidth]{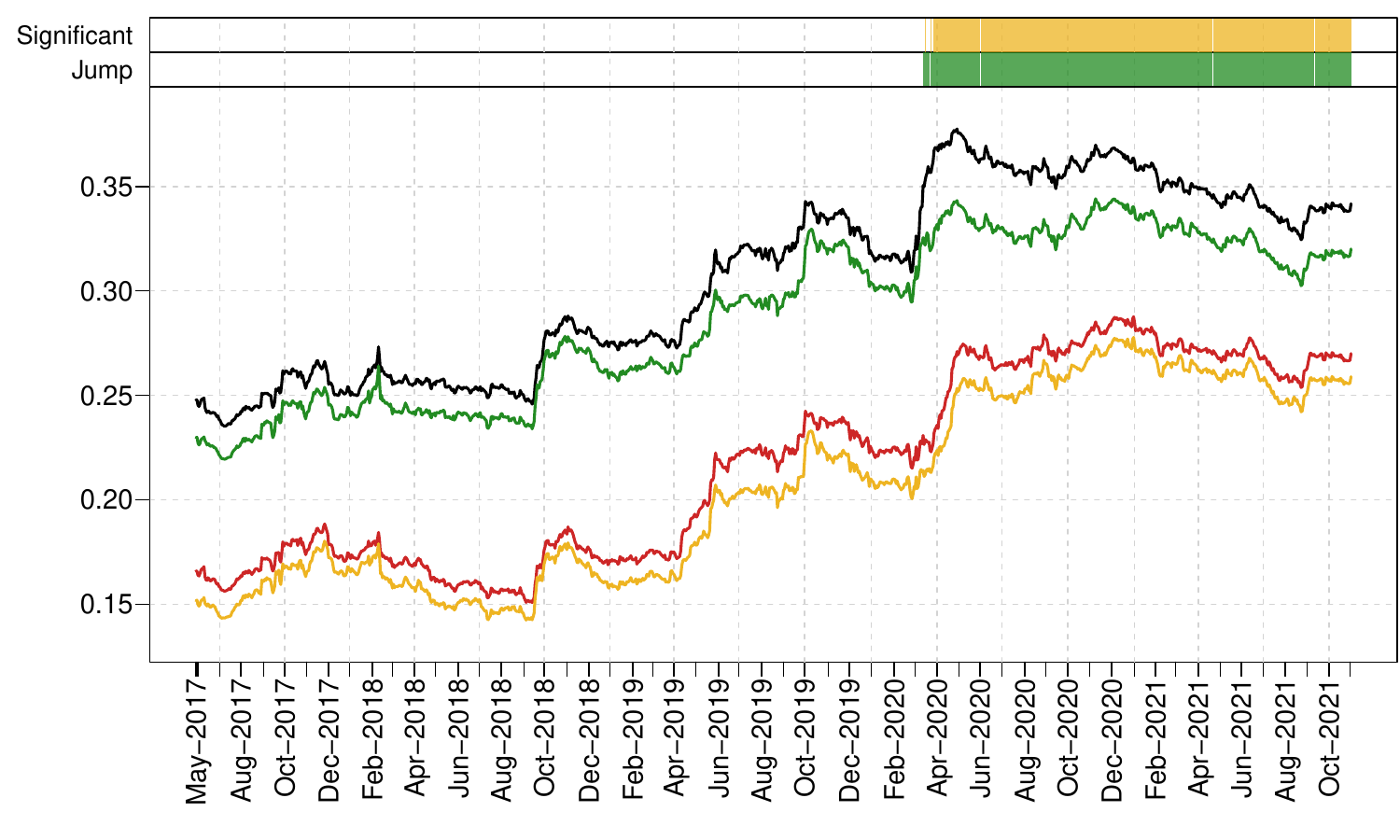}
\caption{$\hat\alpha^{(w)}$}
\end{subfigure}
\hfill
\begin{subfigure}{0.49\textwidth}
\includegraphics[width=\linewidth]{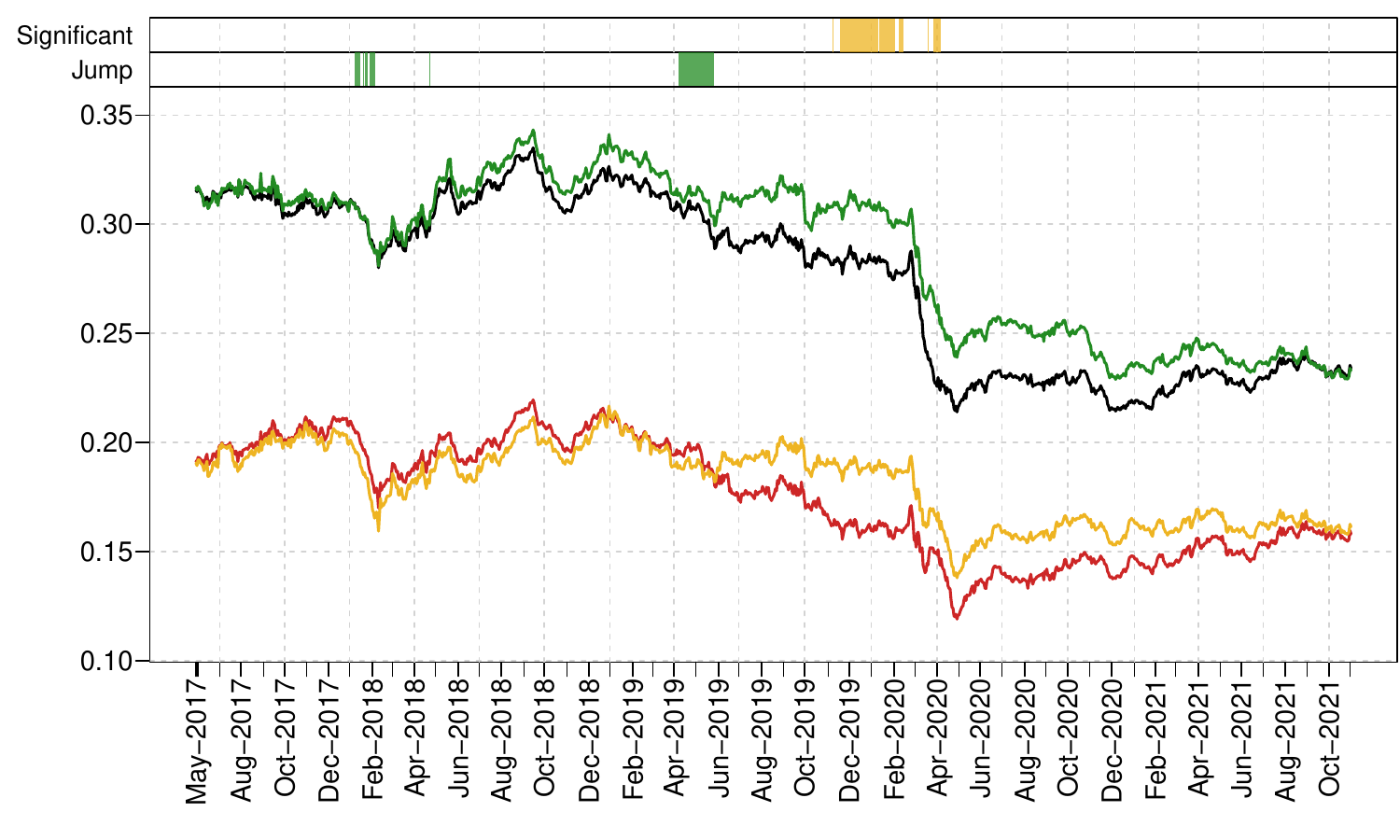}
\caption{$\hat\alpha^{(m)}$}
\end{subfigure}
\vspace{0.5cm}

\begin{subfigure}{0.49\textwidth}
\includegraphics[width=\linewidth]{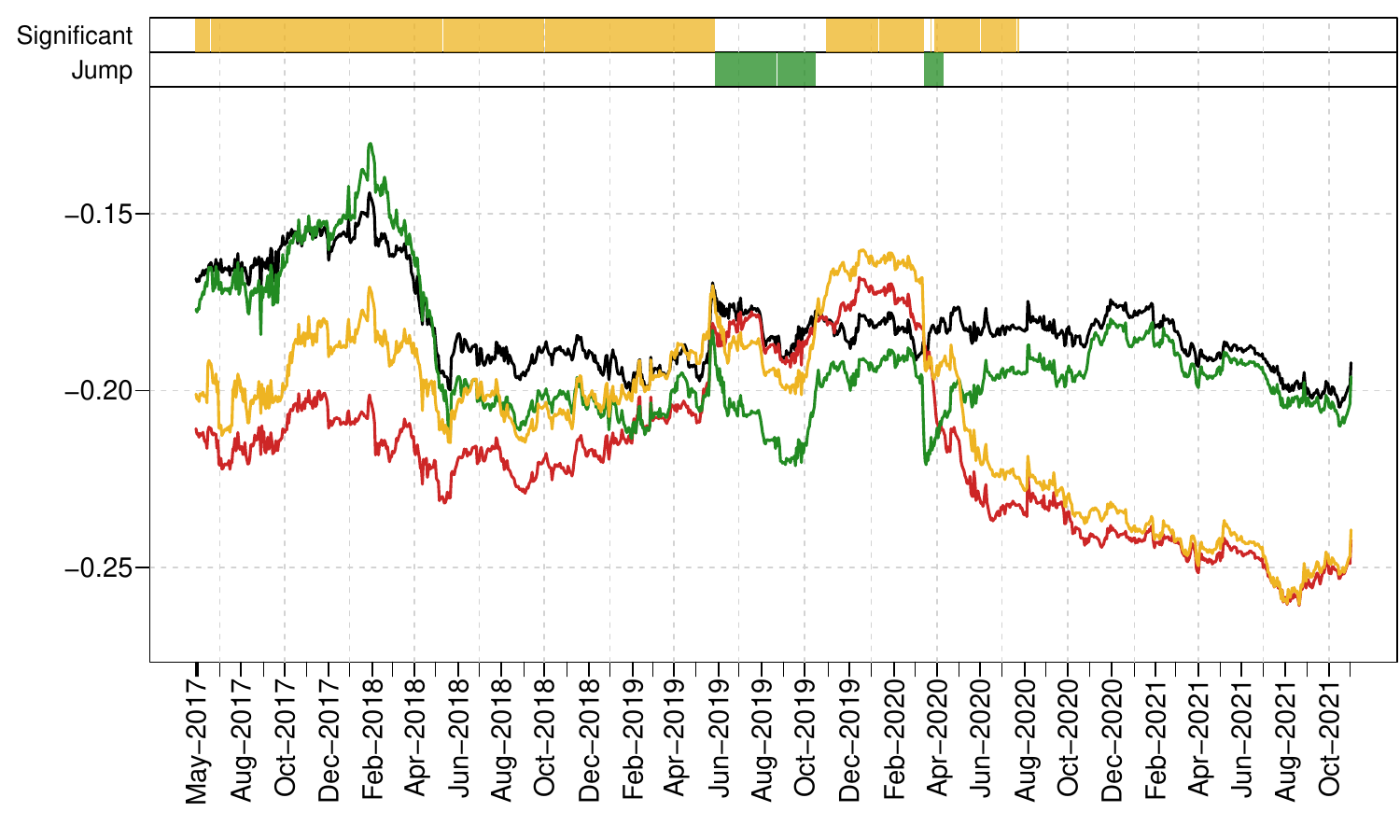}
\caption{$\hat\beta_{1}^{(m)}$}
\end{subfigure}
\hfill
\begin{subfigure}{0.49\textwidth}
\includegraphics[width=\linewidth]{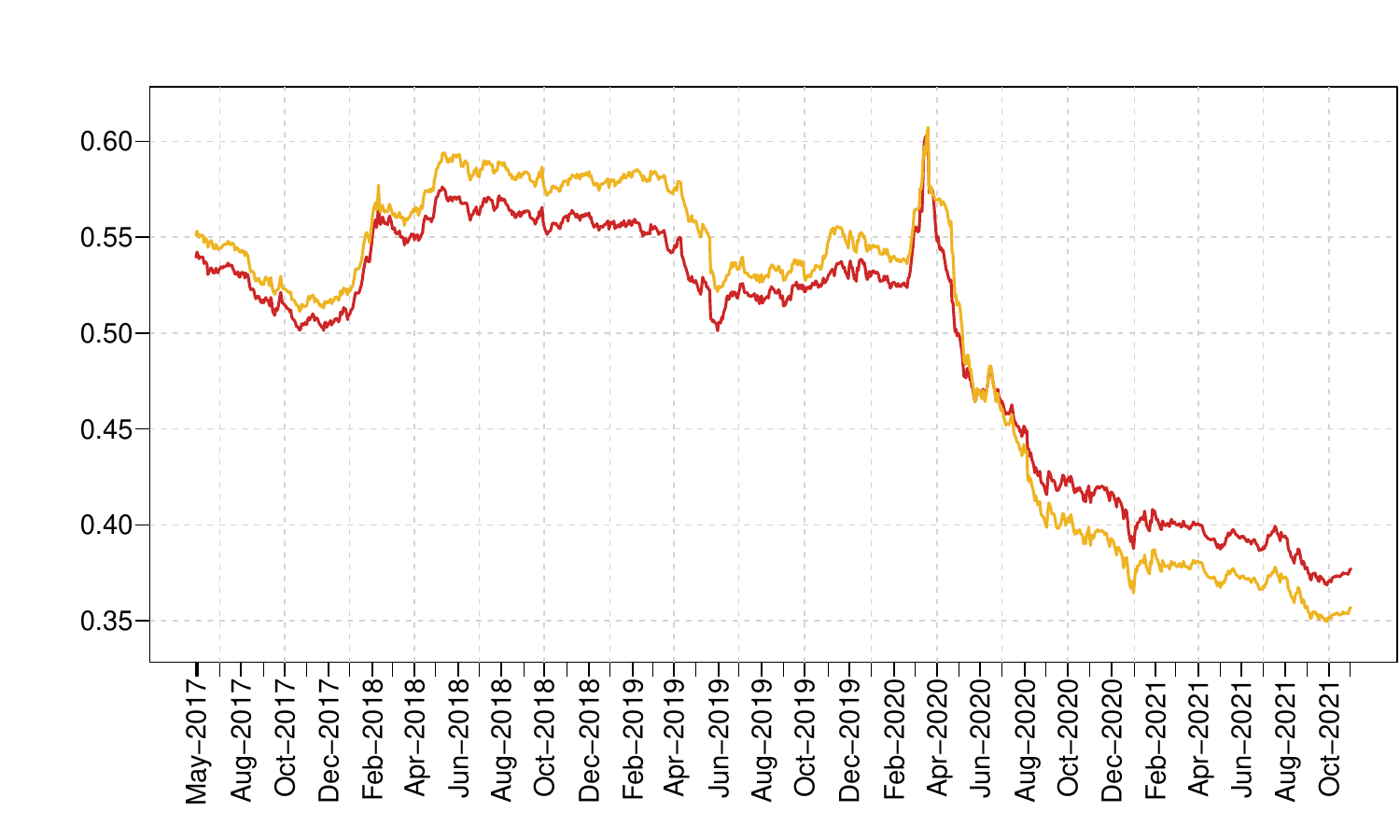}
\caption{$\hat\lambda_{1}$}
\end{subfigure}

\caption{Parameter estimates over the CI22 graph for the global-$\alpha$ GNHAR(1,0,1) model (black lines), the GNHARX([1,0,1],1) model (red lines), and the continuous components of both the GNHAR(1,0,1) model (green lines) and the GNHARX([1,0,1],1) model (yellow lines). Shaded regions indicate where the corresponding jump estimate is statistically significant.} \label{fig:GNHAR101-GNHARX1011-JC-estimates}
\end{figure}

Beyond predictive accuracy, the parsimony of the proposed network-based framework permits a direct examination of the model parameters. By tracking the evolution of estimated coefficients over time, we can identify structural breaks and regime shifts in volatility spillovers that static point-estimates inherently obscure. This provides a quantitative and more granular narrative, facilitating the interpretation of global economic dynamics.
In this section, we examine the estimated coefficients of the global-$\alpha$ GNHAR(1,0,1) model based on the CI22 graph, alongside those of the global-$\{\alpha,\lambda\}$ GNHARX([1,0,1],1) model. We also report the estimates of the continuous component of global-$\alpha$ JC-GNHAR(1,0,1) and global-$\{\alpha,\lambda\}$ JC-GNHARX([1,0,1],1) model. \cref{fig:GNHAR101-GNHARX1011-JC-estimates} presents these time-varying estimated parameters, where the displayed dates correspond to the final day of each rolling window. At the top of each figure, we highlight periods where the null hypothesis of a zero jump is rejected at the $5\%$ significance level.

We see the estimated parameters of all models follow similar patterns throughout the considered period, instantly recalibrating to fit the changing periods. Following a low-volatility regime in 2017, during which all estimated coefficients remained relatively stable, initial fluctuations occur in February 2018, corresponding to the Volmageddon event. Similar financial instability can be observed in the models coefficients throughout 2018, culminating in the market sell-off in the fourth quarter. Finally, significant shifts begin in January 2020, leading to the substantial shock of the COVID-19 pandemic in March 2020, as global economies implemented unprecedented lockdown measures that severely restricted international commerce.

The estimates of the GNHAR and JC-GNHAR models exhibit reasonably similar magnitudes, and the same holds for the GNHARX and JC-GNHARX cases. The jump parameters corresponding to $\hat{\alpha}^{(d)}$ (\cref{fig:GNHAR101-GNHARX1011-JC-estimates}(a)) and $\hat{\alpha}^{(m)}$ (\cref{fig:GNHAR101-GNHARX1011-JC-estimates}(d)) are generally not significantly different from zero, indicating that the daily and monthly continuous components account for most of the dynamics of asset RV. On the other hand, the jump components of the daily network parameter, $\hat{\beta}^{(d)}$ in \cref{fig:GNHAR101-GNHARX1011-JC-estimates}(b), and of the standard weekly parameter, $\hat{\alpha}^{(w)}$ in \cref{fig:GNHAR101-GNHARX1011-JC-estimates}(c), become relevant following the outbreak of the COVID-19 pandemic. In this case, we also notice a more pronounced difference in magnitude between the estimates of the GNHAR and JC-GNHAR models than between those of the exogenous framework.

With the exception of $\hat{\beta}^{(m)}$, the coefficients of the exogenous models exhibit smaller values than those of the standard models, suggesting that the inclusion of the exogenous IV process absorbs a portion of the explanatory power previously captured by the GNHAR lags.
Regarding $\hat{\beta}^{(m)}$ showed in \cref{fig:GNHAR101-GNHARX1011-JC-estimates}(e), it is the only parameter to exhibit negative values and suggest a partial mean-reverting spillover effects. Once the primary trend has been accounted for, the long-term volatility in neighbouring assets may act to suppress volatility in the target asset. This phenomenon is more pronounced in the GNHARX specifications, where the inclusion of the IV process results in greater variability in the $\hat{\beta}^{(m)}$ estimates compared to the other coefficients. Note also that the jump component of $\hat{\beta}^{(m)}$ is estimated to be significantly different from zero depending on whether the exogenous IV regressor is included or not, suggesting that this parameter is quite sensitive to the modelling approach.

By inspecting the plot of $\hat{\lambda}$ (\cref{fig:GNHAR101-GNHARX1011-JC-estimates}(f)), the importance of the exogenous IV process spikes during the onset of the COVID-19 crisis, reflecting the fact that option pricing rapidly assimilated the extreme uncertainty of the global economic shutdown. This peak is followed by a steady decline in $\hat{\lambda}$ over the subsequent months, reflecting that, as global central banks deployed substantial interventions to support the financial system, the model gradually reduces its reliance on IV as a predictor. 
This period is also characterised by an increase of the magnitudes of $\hat{\alpha}^{(d)}$, $\hat{\beta}^{(d)}$, and $\hat{\alpha}^{(w)}$ (\cref{fig:GNHAR101-GNHARX1011-JC-estimates}(a)--(c), respectively), and a decrease in the monthly component $\hat{\alpha}^{(m)}$ (\cref{fig:GNHAR101-GNHARX1011-JC-estimates}(d)). This suggests that daily and weekly dynamics became more significant than long-term trends, as is expected during periods of unprecedented crisis.
Moreover, the spike in daily spillover $\hat{\beta}^{(d)}$ is indicative of severe market contagion, where assets become highly correlated and a shock in one sector rapidly spills over into others within a single trading day.
Concerning the post-2020 period, both $\hat{\alpha}^{(\cdot)}$ and $\hat{\beta}^{(\cdot)}$ estimates seems to return to configurations characterised by smaller fluctuations, reflecting a more stable and persistent system similar to the pre-pandemic period.

\section{Conclusion}\label{sec:concs}
In this article, we proposed a network time series framework combined with HAR model to forecast the log-transformed realized variances of ten international stock indices. Based on five distinct graphs, we evaluated various model configurations, encompassing different constraints on the model coefficients and network interactions across multiple HAR components. We also proposed a model variant which incorporates a continuous-jump decomposition and integrated an option-implied volatility process within the graphical representation.

Our out-of-sample forecasting analysis demonstrated that incorporating network structure was not just a marginal enhancement, rather it was essential for significantly improving the predictive accuracy with this dataset. In addition to relying on individual asset's historical time series, our proposed GNHAR models were able to fully exploit cross-sectional dependencies, thereby enhancing long-term forecasting performance. Furthermore, capturing a global dependency structure provided more stable forecasts than modelling individual asset autoregressions. When the model architecture is correctly specified, the jump-continuous decomposition produced accurate predictions. However, the JC-GNHAR models also exhibited greater sensitivity than the standard GNHAR approach, which remained more robust to specification.

Our findings also highlighted the importance of monitoring cross-market contagion through short-term network parameters. Assets that appear uncorrelated during tranquil periods become highly interdependent during financial crises, so the daily network parameter facilitates accounting for rapid spillover contagion. Finally, rolling-window parameter estimates provided empirical evidence against the use of static volatility models in risk management and served as quantitative indicators of systemic risk. They also illustrated a clear shift from long-term trends to short-term reactions during periods of crisis.

\paragraph{Acknowledgements.} We would like to thank to Marcos Tapia Costa for the useful discussion when preparing this manuscript, and Almut Veraart for providing the RV data used in this article, shared with permission from Mikko Pakkanen. Original data source: Oxford Man Institute (\url{https://oxford-man.ox.ac.uk/research/realized-library}). 

\paragraph{Funding.} CB is supported by a scholarship from the EPSRC Centre for Doctoral Training in Statistical Applied Mathematics at Bath (SAMBa), under the project EP/S022945/1; MAN gratefully acknowledges support from EPSRC grant EP/X002195/1.

\newpage
\bibliographystyle{agsm}
{\fontsize{12pt}{13pt}\selectfont \bibliography{GNHAR-references.bib}}

@Manual{MCS,
    title = {{MCS}: Model Confidence Set Procedure},
    author = {Leopoldo Catania},
    year = {2026},
    note = {R package version 0.2},
    url = {https://CRAN.R-project.org/package=MCS }
}

@Manual{forecast,
    title = {{forecast}: Forecasting functions for time series and
      linear models},
    author = {Rob Hyndman and George Athanasopoulos and Christoph
      Bergmeir and Gabriel Caceres and Leanne Chhay and Mitchell
      O'Hara-Wild and Fotios Petropoulos and Slava Razbash and Earo
      Wang and Farah Yasmeen},
    year = {2026},
    note = {R package version 9.0.2},
    url = {https://pkg.robjhyndman.com/forecast/},
    doi = {10.32614/CRAN.package.forecast},
}

@article{andersen2003modeling,
  title={Modeling and forecasting realized volatility},
  author={Andersen, Torben G and Bollerslev, Tim and Diebold, Francis X and Labys, Paul},
  journal={Econometrica},
  volume={71},
  number={2},
  pages={579--625},
  year={2003},
  publisher={Wiley Online Library}
}

@article{andersen2007roughing,
  title={Roughing it up: {I}ncluding jump components in the measurement, modeling, and forecasting of return volatility},
  author={Andersen, Torben G and Bollerslev, Tim and Diebold, Francis X},
  journal={The Review of Economics and Statistics},
  volume={89},
  number={4},
  pages={701--720},
  year={2007},
  publisher={The MIT Press}
}

@article{audrino2020impact,
  title={The impact of sentiment and attention measures on stock market volatility},
  author={Audrino, Francesco and Sigrist, Fabio and Ballinari, Daniele},
  journal={International Journal of Forecasting},
  volume={36},
  number={2},
  pages={334--357},
  year={2020},
  publisher={Elsevier}
}

@article{bandi2006separating,
  title={Separating microstructure noise from volatility},
  author={Bandi, Federico M and Russell, Jeffrey R},
  journal={Journal of Financial Economics},
  volume={79},
  number={3},
  pages={655--692},
  year={2006},
  publisher={Elsevier}
}

@article{barndorff2002econometric,
  title={Econometric analysis of realized volatility and its use in estimating stochastic volatility models},
  author={Barndorff-Nielsen, Ole E and Shephard, Neil},
  journal={Journal of the Royal Statistical Society Series B: Statistical Methodology},
  volume={64},
  number={2},
  pages={253--280},
  year={2002},
  publisher={Oxford University Press}
}

@article{bekaert2014vix,
  title={The {VIX}, the variance premium and stock market volatility},
  author={Bekaert, Geert and Hoerova, Marie},
  journal={Journal of Econometrics},
  volume={183},
  number={2},
  pages={181--192},
  year={2014},
  publisher={Elsevier}
}

@article{blair2001forecasting,
  title={Forecasting {S}\&{P} 100 volatility: the incremental information content of implied volatilities and high-frequency index returns},
  author={Blair, Bevan J and Poon, Ser-Huang and Taylor, Stephen J},
  journal={Journal of Econometrics},
  volume={105},
  number={1},
  pages={5--26},
  year={2001},
  publisher={Elsevier}
}

@article{boetti2025long,
  title={Long memory network time series},
  author={Boetti, Chiara and Nunes, Matthew A and Knight, Marina I},
  journal={arXiv preprint arXiv:2512.10446},
  year={2025}
}

@article{bollerslev1986generalized,
  title={Generalized autoregressive conditional heteroskedasticity},
  author={Bollerslev, Tim},
  journal={Journal of Econometrics},
  volume={31},
  number={3},
  pages={307--327},
  year={1986},
  publisher={Elsevier}
}

@article{britten2000option,
  title={Option prices, implied price processes, and stochastic volatility},
  author={Britten-Jones, Mark and Neuberger, Anthony},
  journal={Journal of Finance},
  volume={55},
  number={2},
  pages={839--866},
  year={2000},
  publisher={Wiley Online Library}
}

@article{bubak2011volatility,
  title={Volatility transmission in emerging {European} foreign exchange markets},
  author={Bub{\'a}k, V{\'\i}t and Ko{\v{c}}enda, Ev{\v{z}}en and {\v{Z}}ike{\v{s}}, Filip},
  journal={Journal of Banking \& Finance},
  volume={35},
  number={11},
  pages={2829--2841},
  year={2011},
  publisher={Elsevier}
}

@article{busch2011role,
  title={The role of implied volatility in forecasting future realized volatility and jumps in foreign exchange, stock, and bond markets},
  author={Busch, Thomas and Christensen, Bent Jesper and Nielsen, Morten {\O}rregaard},
  journal={Journal of Econometrics},
  volume={160},
  number={1},
  pages={48--57},
  year={2011},
  publisher={Elsevier}
}

@article{chiriac2011modelling,
  title={Modelling and forecasting multivariate realized volatility},
  author={Chiriac, Roxana and Voev, Valeri},
  journal={Journal of Applied Econometrics},
  volume={26},
  number={6},
  pages={922--947},
  year={2011},
  publisher={Wiley Online Library}
}

@article{christensen1998relation,
  title={The relation between implied and realized volatility},
  author={Christensen, Bent J and Prabhala, Nagpurnanand R},
  journal={Journal of Financial Economics},
  volume={50},
  number={2},
  pages={125--150},
  year={1998},
  publisher={Elsevier}
}

@article{corsi2009simple,
  title={A simple approximate long-memory model of realized volatility},
  author={Corsi, Fulvio},
  journal={Journal of Financial Econometrics},
  volume={7},
  number={2},
  pages={174--196},
  year={2009},
  publisher={Oxford University Press}
}

@article{cubadda2017vector,
  title={A vector heterogeneous autoregressive index model for realized volatility measures},
  author={Cubadda, Gianluca and Guardabascio, Barbara and Hecq, Alain},
  journal={International Journal of Forecasting},
  volume={33},
  number={2},
  pages={337--344},
  year={2017},
  publisher={Elsevier}
}

@article{diebold1995comparing,
  title={Comparing predictive accuracy},
  author={Diebold, Francis X and Mariano, Robert S},
  journal={Journal of Business \& Economic Statistics},
  volume={13},
  number={3},
  pages={134--144},
  year={1995},
  publisher={Taylor \& Francis}
}

@article{diebold2009measuring,
  title={Measuring financial asset return and volatility spillovers, with application to global equity markets},
  author={Diebold, Francis X and Yilmaz, Kamil},
  journal={The Economic Journal},
  volume={119},
  number={534},
  pages={158--171},
  year={2009},
  publisher={Oxford University Press Oxford, UK}
}

@article{diebold2012better,
  title={Better to give than to receive: {Predictive} directional measurement of volatility spillovers},
  author={Diebold, Francis X and Yilmaz, Kamil},
  journal={International Journal of Forecasting},
  volume={28},
  number={1},
  pages={57--66},
  year={2012},
  publisher={Elsevier}
}

@article{diebold2014network,
  title={On the network topology of variance decompositions: {Measuring} the connectedness of financial firms},
  author={Diebold, Francis X and Y{\i}lmaz, Kamil},
  journal={Journal of Econometrics},
  volume={182},
  number={1},
  pages={119--134},
  year={2014},
  publisher={Elsevier}
}

@article{engle1982autoregressive,
  title={Autoregressive conditional heteroscedasticity with estimates of the variance of {United Kingdom} inflation},
  author={Engle, Robert F},
  journal={Econometrica},
  pages={987--1007},
  year={1982},
  publisher={JSTOR}
}

@article{gneiting2011making,
  title={Making and evaluating point forecasts},
  author={Gneiting, Tilmann},
  journal={Journal of the American Statistical Association},
  volume={106},
  number={494},
  pages={746--762},
  year={2011},
  publisher={Taylor \& Francis}
}

@article{granger1969investigating,
  title={Investigating causal relations by econometric models and cross-spectral methods},
  author={Granger, Clive WJ},
  journal={Econometrica},
  pages={424--438},
  year={1969},
  publisher={JSTOR}
}

@article{hansen2006realized,
  title={Realized variance and market microstructure noise},
  author={Hansen, Peter R and Lunde, Asger},
  journal={Journal of Business \& Economic Statistics},
  volume={24},
  number={2},
  pages={127--161},
  year={2006},
  publisher={Taylor \& Francis}
}

@article{hansen2011model,
  title={The model confidence set},
  author={Hansen, Peter R and Lunde, Asger and Nason, James M},
  journal={Econometrica},
  volume={79},
  number={2},
  pages={453--497},
  year={2011},
  publisher={Wiley Online Library}
}

@article{harvey1997testing,
  title={Testing the equality of prediction mean squared errors},
  author={Harvey, David and Leybourne, Stephen and Newbold, Paul},
  journal={International Journal of Forecasting},
  volume={13},
  number={2},
  pages={281--291},
  year={1997},
  publisher={Elsevier}
}

@article{knight2016arxiv,
   author = {{Knight}, M.~I. and {Nunes}, M.~A. and {Nason}, G.~P.},
    title = "{Modelling, Detrending and Decorrelation of Network Time Series}",
  journal = {ArXiv e-prints},
archivePrefix = "arXiv",
   volume = {1603.03221},
 primaryClass = "stat.ME",
     year = 2016
}

@article{knight2020generalized,
  title={Generalized network autoregressive processes and the {GNAR} package},
  author={Knight, Marina and Leeming, Kathryn and Nason, Guy and Nunes, Matthew},
  journal={Journal of Statistical Software},
  volume={96},
  pages={1--36},
  year={2020}
}

@article{liang2020implied,
  title={Is implied volatility more informative for forecasting realized volatility: {A}n international perspective},
  author={Liang, Chao and Wei, Yu and Zhang, Yaojie},
  journal={Journal of Forecasting},
  volume={39},
  number={8},
  pages={1253--1276},
  year={2020},
  publisher={Wiley Online Library}
}

@article{liu2015economic,
  title={Economic policy uncertainty and stock market volatility},
  author={Liu, Li and Zhang, Tao},
  journal={Finance Research Letters},
  volume={15},
  pages={99--105},
  year={2015},
  publisher={Elsevier}
}

@article{luo2025forecasting,
  title={Forecasting multivariate volatilities with exogenous predictors: An application to industry diversification strategies},
  author={Luo, Jiawen and Cepni, Oguzhan and Demirer, Riza and Gupta, Rangan},
  journal={Journal of Empirical Finance},
  volume={81},
  pages={101595},
  year={2025},
  publisher={Elsevier}
}

@article{marcellino2006comparison,
  title={A comparison of direct and iterated multistep {AR} methods for forecasting macroeconomic time series},
  author={Marcellino, Massimiliano and Stock, James H and Watson, Mark W},
  journal={Journal of Econometrics},
  volume={135},
  number={1-2},
  pages={499--526},
  year={2006},
  publisher={Elsevier}
}

@article{nason2013test,
  title={A test for second-order stationarity and approximate confidence intervals for localized autocovariances for locally stationary time series},
  author={Nason, Guy},
  journal={Journal of the Royal Statistical Society Series B},
  volume={75},
  number={5},
  pages={879--904},
  year={2013},
  publisher={Oxford University Press}
}

@article{nason2022quantifying,
  title={Quantifying the economic response to {COVID}-19 mitigations and death rates via forecasting purchasing managers' indices using generalised network autoregressive models with exogenous variables},
  author={Nason, Guy P and Wei, James L},
  journal={Journal of the Royal Statistical Society Series A},
  volume={185},
  number={4},
  pages={1778--1792},
  year={2022},
  publisher={Oxford University Press}
}

@article{nason2025forecasting,
  title={Forecasting {UK} consumer price inflation with {RaGNAR}: {Random} generalised network autoregressive processes},
  author={Nason, Guy P and Palasciano, Henry Antonio},
  journal={International Journal of Forecasting},
  year={2025},
  publisher={Elsevier}
}

@inproceedings{nason2024modelling,
  title={Modelling clusters in network time series with an application to presidential elections in the {USA}},
  author={Nason, Guy and Salnikov, Daniel and Cortina-Borja, Mario},
  booktitle={Conference of the International Federation of Classification Societies},
  pages={115--123},
  year={2024},
  organization={Springer}
}

@article{oh2016high,
  title={High-dimensional copula-based distributions with mixed frequency data},
  author={Oh, Dong Hwan and Patton, Andrew J},
  journal={Journal of Econometrics},
  volume={193},
  number={2},
  pages={349--366},
  year={2016},
  publisher={Elsevier}
}

@article{poon2003forecasting,
  title={Forecasting volatility in financial markets: A review},
  author={Poon, Ser-Huang and Granger, Clive W J},
  journal={Journal of Economic Literature},
  volume={41},
  number={2},
  pages={478--539},
  year={2003},
  publisher={American Economic Association}
}

@article{shi2020comparison,
  title={A comparison of conditional predictive ability of implied volatility and realized measures in forecasting volatility},
  author={Shi, Yafeng and Ying, Tingting and Shi, Yanlong and Ai, Chunrong},
  journal={Journal of Forecasting},
  volume={39},
  number={7},
  pages={1025--1034},
  year={2020},
  publisher={Wiley Online Library}
}

@article{son2023forecasting,
  title={Forecasting global stock market volatility: the impact of volatility spillover index in spatial-temporal graph-based model},
  author={Son, Bumho and Lee, Yunyoung and Park, Seongwan and Lee, Jaewook},
  journal={Journal of Forecasting},
  volume={42},
  number={7},
  pages={1539--1559},
  year={2023},
  publisher={Wiley Online Library}
}

@article{tapia2025higher,
  title={Higher Order Dynamic Network Linear Models for Covariance Forecasting},
  author={Tapia Costa, Marcos and Cucuringu, Mihai and Nason, Guy P},
  journal={Available at SSRN 5113698},
  year={2025}
}

@article{taylor1994modeling,
  title={Modeling stochastic volatility: {A} review and comparative study},
  author={Taylor, Stephen J},
  journal={Mathematical Finance},
  volume={4},
  number={2},
  pages={183--204},
  year={1994},
  publisher={Wiley Online Library}
}

@article{taylor2015realized,
  title={Realized volatility forecasting in an international context},
  author={Taylor, Nicholas},
  journal={Applied Economics Letters},
  volume={22},
  number={6},
  pages={503--509},
  year={2015},
  publisher={Taylor \& Francis}
}

@article{wen2019forecasting,
  title={Forecasting realized volatility of crude oil futures with equity market uncertainty},
  author={Wen, Fenghua and Zhao, Yupei and Zhang, Minzhi and Hu, Chunyan},
  journal={Applied Economics},
  volume={51},
  number={59},
  pages={6411--6427},
  year={2019},
  publisher={Taylor \& Francis}
}

@article{wilms2021multivariate,
  title={Multivariate volatility forecasts for stock market indices},
  author={Wilms, Ines and Rombouts, Jeroen and Croux, Christophe},
  journal={International Journal of Forecasting},
  volume={37},
  number={2},
  pages={484--499},
  year={2021},
  publisher={Elsevier}
}

@article{zhang2025graph,
  title={Graph-based methods for forecasting realized covariances},
  author={Zhang, Chao and Pu, Xingyue and Cucuringu, Mihai and Dong, Xiaowen},
  journal={Journal of Financial Econometrics},
  volume={23},
  number={2},
  pages={nbae026},
  year={2025},
  publisher={Oxford University Press}
}

@article{zhang2025forecasting,
  title={Forecasting realized volatility with spillover effects: {P}erspectives from graph neural networks},
  author={Zhang, Chao and Pu, Xingyue and Cucuringu, Mihai and Dong, Xiaowen},
  journal={International Journal of Forecasting},
  volume={41},
  number={1},
  pages={377--397},
  year={2025},
  publisher={Elsevier}
}

@article{zhu2017network,
    author = {Zhu, Xuening and Pan, Rui and Li, Guodong and Liu, Yuewen and Wang, Hansheng},
title = {Network vector autoregression},
    journal = {The Annals of Statistics},
    number = {3},
    pages = {1096-1123},
    volume = {45},
    year = {2017}
}

\appendix
\setcounter{section}{0}
\setcounter{figure}{0}
\renewcommand{\thefigure}{\thesection.\arabic{figure}}
\setcounter{table}{0}
\renewcommand{\thetable}{\thesection.\arabic{table}}

\section{Additional Results}\label{app:additional}
This appendix provides the additional results and analyses referenced in the main text.

\subsection{Data Summaries}
\cref{tab:summary-stats-logRV} and \cref{tab:summary-stats-2logIV} report summary statistics for the log-realized and log-implied variance time series, respectively, including the mean, standard deviation, skewness, kurtosis, first-order autocorrelation, and percentage of missing data. Tables also present the number of rejections for \citet{nason2013test}'s second-order stationarity test, once missing values have been linearly interpolated. To satisfy the test's requirement for dyadic-length series, we split the sample into two overlapping periods of length 1024 observations. The first period runs from August 6, 2013 to July 3, 2017, and the second from February 2, 2018 to January 3, 2022. Note the latter includes COVID-19, thus we expect the processes to be non-stationary. 

\begin{table}[H]
\centering
{\setlength{\tabcolsep}{9pt}\small
\begin{tabular}{l|cccccccc}
\toprule
\textbf{Index} & \textbf{Mean} & \textbf{SD} & \textbf{Skew} & \textbf{Kurt} & \textbf{AC1} & $\boldsymbol{\%}$ \textbf{NA} & \textbf{Rej.1} & \textbf{Rej.2} \\
\midrule
DJI & -10.322 & 1.097 & 0.666 & 4.094 & 0.774 & 4.36 & 0 & 11 \\
GDAXI & -9.840 & 0.883 & 0.478 & 3.941 & 0.728 & 3.72 & 1 & 14 \\
HSI & -9.958 & 0.692 & 0.595 & 4.487 & 0.621 & 6.40 & 2 & 3 \\
IXIC & -10.150 & 1.002 & 0.681 & 4.025 & 0.774 & 4.09 & 0 & 6 \\
KS11 & -10.338 & 0.759 & 1.040 & 5.846 & 0.732 & 6.40 & 2 & 14 \\
N225 & -10.117 & 0.925 & 0.699 & 4.185 & 0.711 & 7.18 & 3 & 9 \\
NSEI & -10.106 & 0.815 & 0.901 & 6.051 & 0.677 & 6.22 & 0 & 16 \\
RUT & -10.080 & 0.929 & 0.749 & 4.384 & 0.737 & 4.22 & 2 & 12 \\
SPX & -10.433 & 1.116 & 0.656 & 4.015 & 0.796 & 4.27 & 0 & 6 \\
STOXX50E & -9.746 & 0.992 & -0.095 & 8.026 & 0.672 & 3.00 & 2 & 16 \\
\bottomrule
\end{tabular}}
\caption{Summary statistics for log-realized variance time series.}
\label{tab:summary-stats-logRV}
\end{table}

\begin{table}[H]
\centering
{\setlength{\tabcolsep}{9pt}\small
\begin{tabular}{l|cccccccc}
\toprule
\textbf{Index} & \textbf{Mean} & \textbf{SD} & \textbf{Skew} & \textbf{Kurt} & \textbf{AC1} & $\boldsymbol{\%}$ \textbf{NA} & \textbf{Rej.1} & \textbf{Rej.2} \\
\midrule
DJI & 5.539 & 0.638 & 1.264 & 5.196 & 0.975 & 3.72 & 4 & 17 \\
GDAXI & 5.884 & 0.603 & 0.956 & 4.569 & 0.976 & 3.45 & 0 & 25 \\
HSI & 5.879 & 0.491 & 0.727 & 4.371 & 0.972 & 5.77 & 8 & 20 \\
IXIC & 5.906 & 0.614 & 0.949 & 3.869 & 0.975 & 3.68 & 5 & 17 \\
KS11 & 5.465 & 0.574 & 1.568 & 6.206 & 0.976 & 6.13 & 7 & 23 \\
N225 & 6.078 & 0.514 & 0.673 & 3.724 & 0.968 & 6.72 & 11 & 17 \\
NSEI & 5.665 & 0.568 & 1.488 & 7.001 & 0.984 & 5.77 & 12 & 21 \\
RUT & 6.058 & 0.599 & 1.243 & 4.769 & 0.978 & 3.77 & 6 & 20 \\
SPX & 5.567 & 0.657 & 1.144 & 5.010 & 0.969 & 3.45 & 4 & 17 \\
STOXX50E & 5.894 & 0.596 & 0.865 & 4.579 & 0.970 & 2.82 & 0 & 25 \\
\bottomrule
\end{tabular}}
\caption{Summary statistics for log-implied variance time series.}\label{tab:summary-stats-2logIV}
\end{table}

\subsection{Time Plot Absolute Forecast Error}

\begin{figure}[H]
\centering
\begin{subfigure}{0.65\textwidth}
\includegraphics[width=\linewidth]{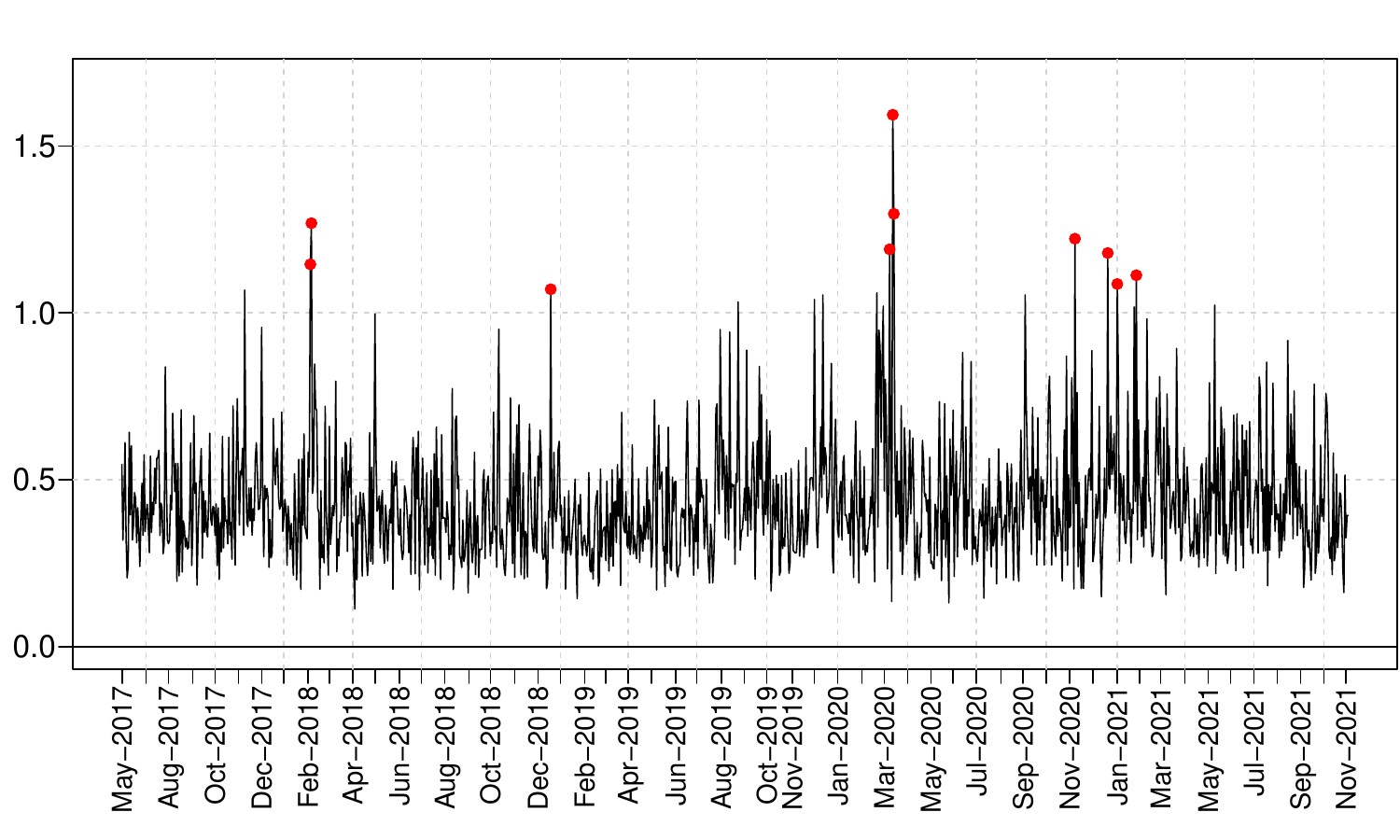}
\caption{$h=1$}
\end{subfigure}
\vspace{0.25cm}

\begin{subfigure}{0.49\textwidth}
\includegraphics[width=\linewidth]{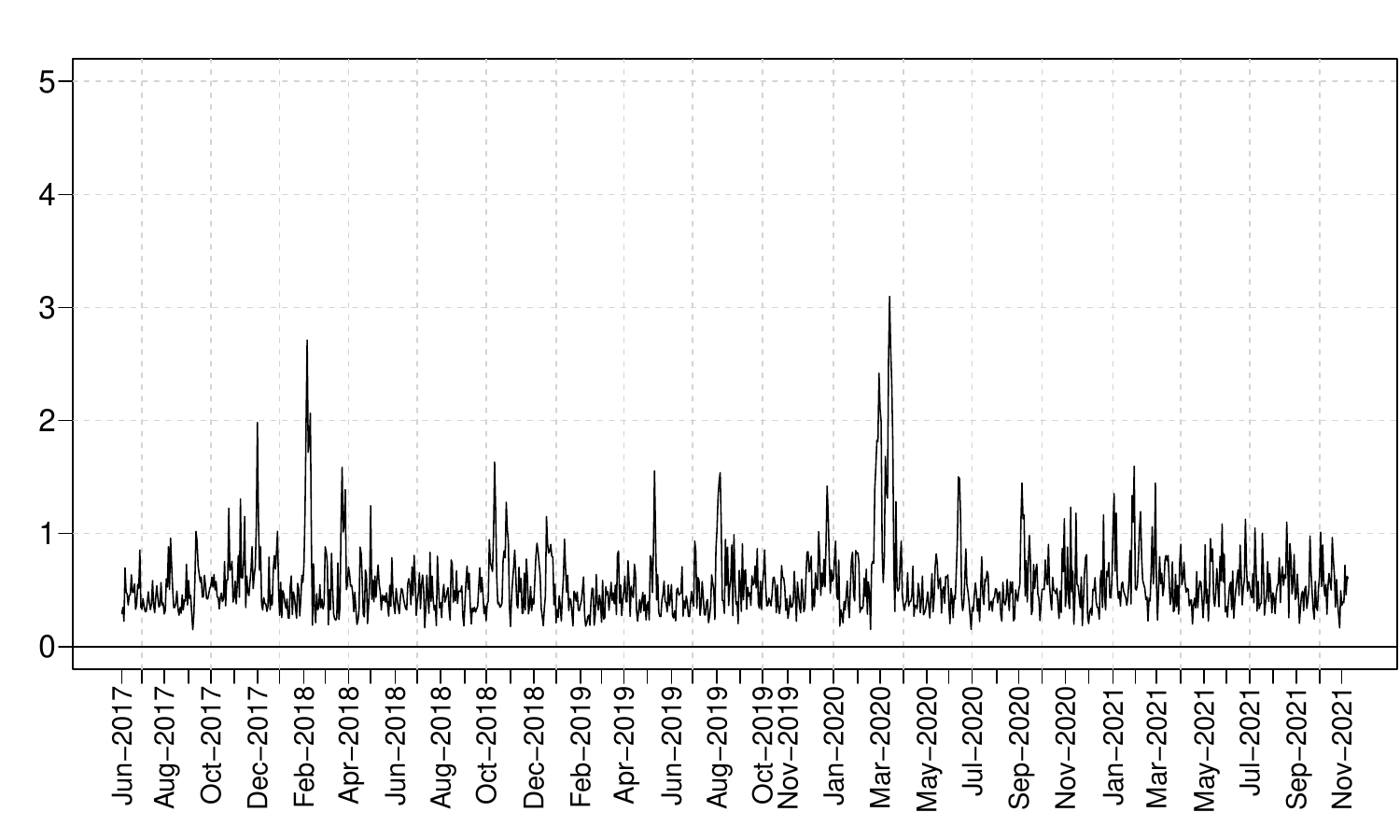}
\caption{$h=5$}
\end{subfigure}
\hfill
\begin{subfigure}{0.49\textwidth}
\includegraphics[width=\linewidth]{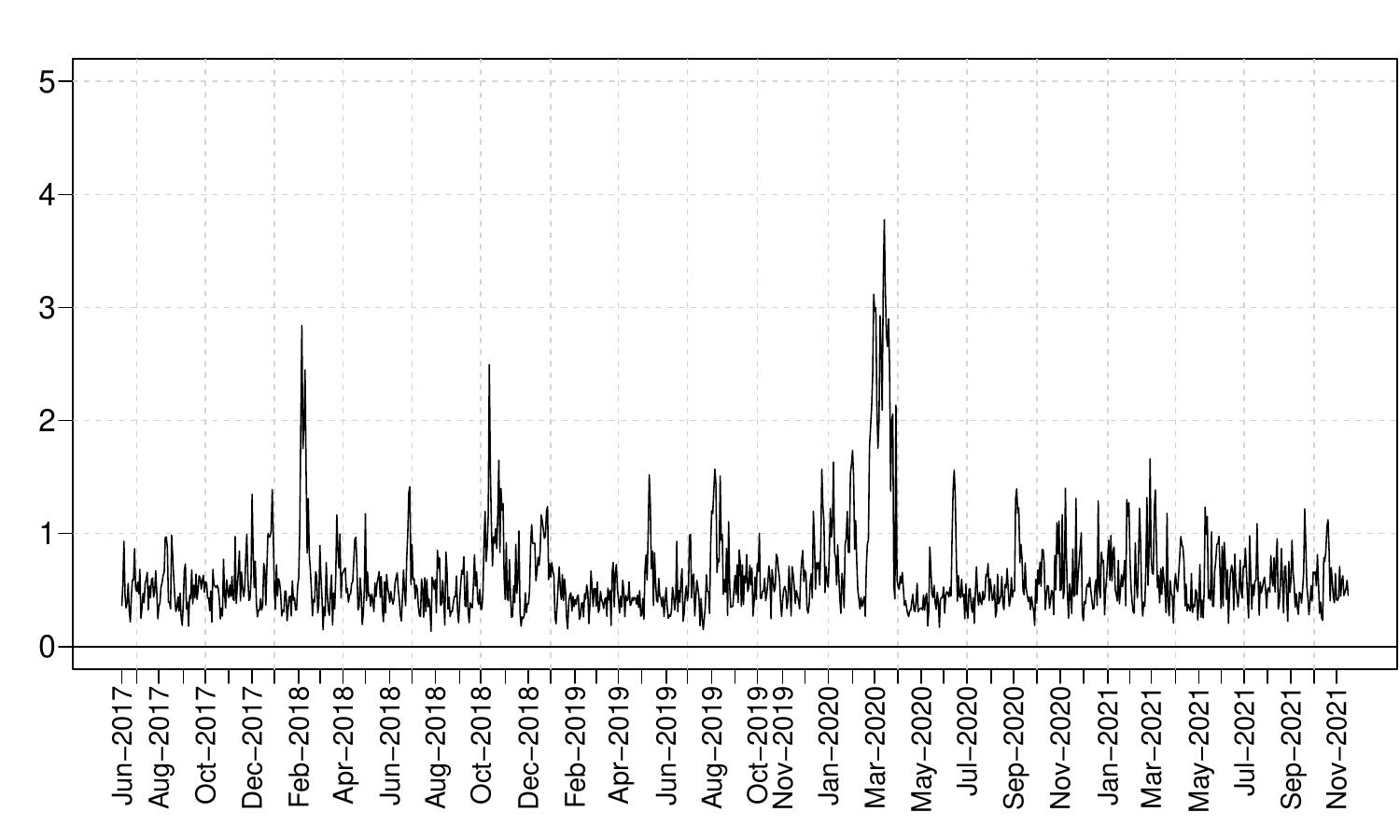}
\caption{$h=10$}
\end{subfigure}
\vspace{0.25cm}

\begin{subfigure}{0.49\textwidth}
\includegraphics[width=\linewidth]{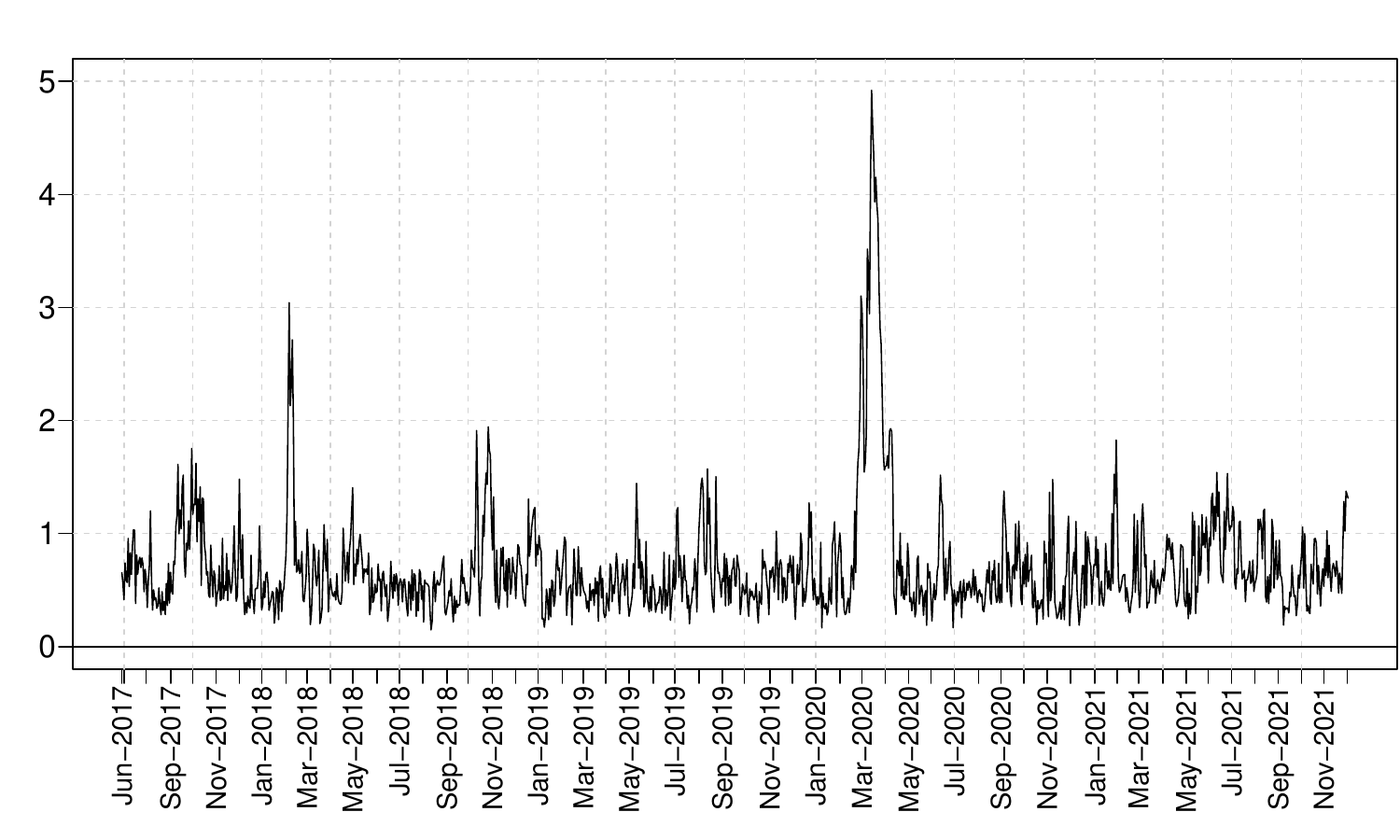}
\caption{$h=22$}
\end{subfigure}
\hfill
\begin{subfigure}{0.49\textwidth}
\includegraphics[width=\linewidth]{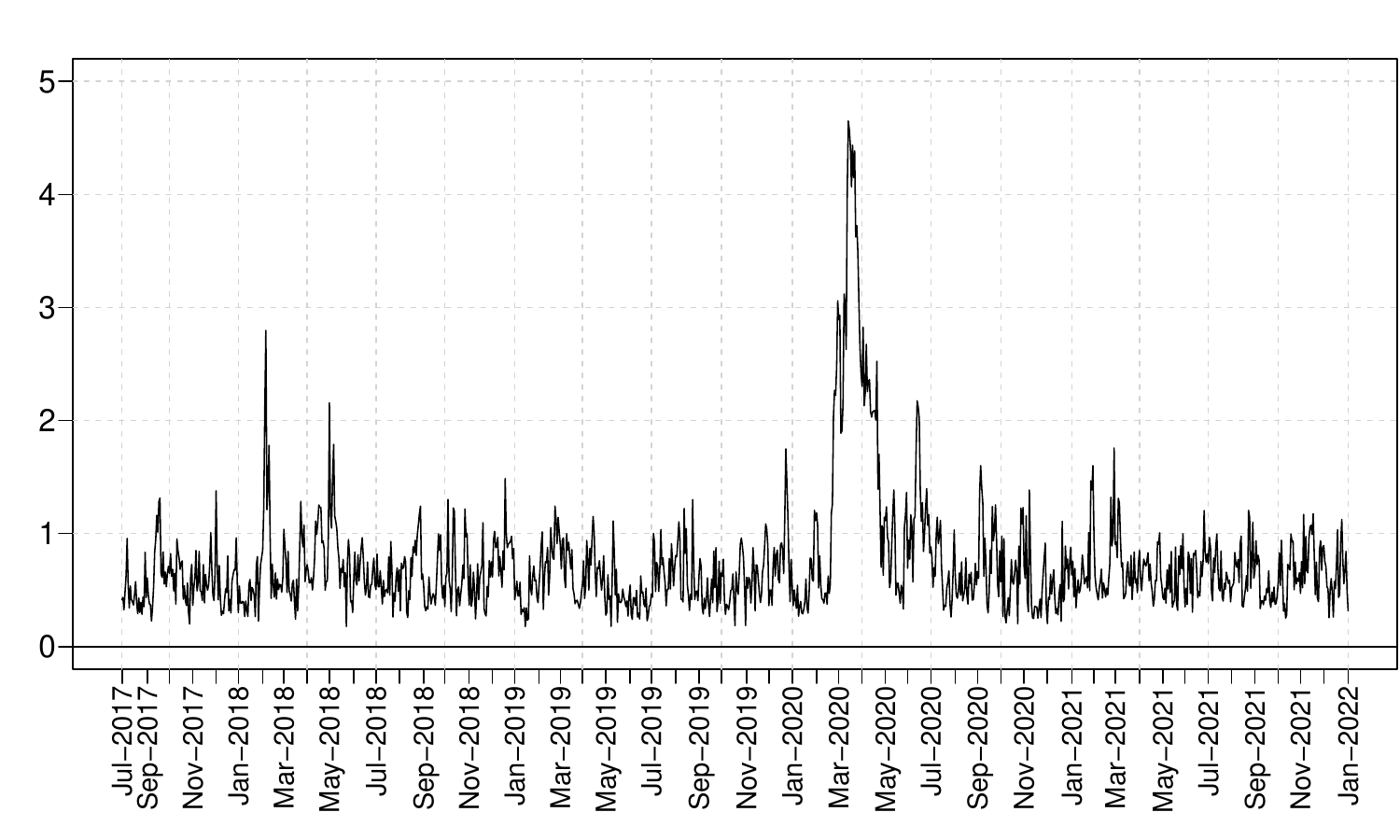}
\caption{$h=44$}
\end{subfigure}

\caption{Time plot of the AFE averaged across the stocks for the global-$\alpha$ JC-GNHAR(1,0,1) model over the CI22 graph. Red dots in (a) indicate the ten largest errors at $h=1$.} \label{fig:JCGNHAR101-MAFE-time}
\end{figure}

\subsection{Tables of Forecast Accuracy with MSFE}

\begin{table}[H]
\centering
{\setlength{\tabcolsep}{9pt}\small
\begin{tabular}{ll|cccc}
\toprule
& & \multicolumn{2}{c}{\textbf{GNHAR}} & \multicolumn{2}{c}{\textbf{JC-GNHAR}} \\
\textbf{Network} & \textbf{Order} & \textbf{gl-}$\boldsymbol{\alpha}$ & \textbf{ind-}$\boldsymbol{\alpha}$ & \textbf{gl-}$\boldsymbol{\alpha}$ & \textbf{ind-}$\boldsymbol{\alpha}$ \\
\midrule
\multirow{7}{*}{Fully} & (1,0,0) & 0.330 (<0.01) & 0.586 (<0.01) & 0.334 (<0.01) & 0.975 (<0.01) \\
& (0,1,0) & 0.333 (<0.01) & 0.608 (<0.01) & 0.330 (<0.01) & 1.175 (<0.01) \\
& (0,0,1) & 0.352 (<0.01) & 0.608 (<0.01) & 0.369 (<0.01) & 1.056 (<0.01) \\
& (0,1,1) & 0.332 (<0.01) & 0.444 (<0.01) & 0.351 (<0.01) & 0.821 (<0.01) \\
& (1,0,1) & \textbf{0.321 (1)} & 0.382 (<0.01)  & 0.340 (<0.01) & 0.518 (<0.01) \\
& (1,1,0) & \textbf{0.322 (0.45)} & 0.485 (<0.01) & 0.327 (0.07) & 1.110 (<0.01) \\
& (1,1,1) & 0.352 (0.01) & 0.420 (<0.01) & 0.347 (<0.01) & 0.707 (<0.01) \\
\midrule
\multirow{7}{*}{GC1} & (1,0,0) & 0.330 (<0.01) & 0.579 (<0.01) & 0.335 (<0.01) & 0.957 (<0.01) \\
& (0,1,0) & 0.333 (<0.01) & 0.591 (<0.01) & 0.330 (<0.01) & 1.148 (<0.01) \\
& (0,0,1) & 0.355 (<0.01) & 0.606 (<0.01) & 0.368 (<0.01) & 1.053 (<0.01) \\
& (0,1,1) & 0.332 (<0.01) & 0.438 (<0.01) & 0.353 (<0.01) & 0.836 (<0.01) \\
& (1,0,1) & \textbf{0.322 (1)} & 0.382 (<0.01) & 0.343 (<0.01) & 0.520 (<0.01) \\
& (1,1,0) & \textbf{0.322 (0.40)} & 0.481 (<0.01) & 0.328 (0.11) & 1.126 (<0.01) \\
& (1,1,1) & 0.363 (0.01) & 0.421 (<0.01) & 0.344 (<0.01) & 0.717 (<0.01) \\ 
\midrule
\multirow{7}{*}{GC22} & (1,0,0) & 0.331 (<0.01) & 0.548 (<0.01) & 0.331 (0.29) & 0.899 (<0.01) \\
& (0,1,0) & 0.334 (<0.01) & 0.498 (<0.01) & 0.332 (0.09) & 0.983 (<0.01) \\
& (0,0,1) & 0.372 (<0.01) & 0.570 (<0.01) & 0.352 (<0.01) & 0.984 (<0.01) \\
& (0,1,1) & 0.363 (<0.01) & 0.407 (<0.01) & 0.354 (<0.01) & 0.692 (<0.01) \\
& (1,0,1) & \textbf{0.326 (1)} & 0.375 (<0.01) & \textbf{0.333 (0.21)} & 0.622 (<0.01) \\
& (1,1,0) & \textbf{0.326 (0.36)} & 0.479 (<0.01) & \textbf{0.330 (0.35)} & 0.952 (<0.01) \\
& (1,1,1) & 0.402 (<0.01) & 0.477 (<0.01) & 0.361 (<0.01) & 0.739 (<0.01) \\
\midrule
\multirow{7}{*}{CI1} & (1,0,0) & 0.332 (<0.01) & 0.553 (<0.01) & 0.328 (0.11) & 0.838 (<0.01) \\
& (0,1,0) & 0.335 (<0.01) & 0.549 (<0.01) & 0.331 (<0.01) & 1.025 (<0.01) \\
& (0,0,1) & 0.364 (<0.01) & 0.657 (<0.01) & 0.354 (<0.01) & 1.209 (<0.01) \\
& (0,1,1) & 0.338 (0.01) & 0.515 (<0.01) & 0.339 (0.11) & 0.832 (<0.01) \\
& (1,0,1) & 0.329 (0.09) & 0.369 (<0.01) & \textbf{0.325 (1)} & 0.586 (<0.01) \\
& (1,1,0) & 0.330 (0.08) & 0.527 (<0.01) & \textbf{0.325 (0.91)} & 0.934 (<0.01) \\
& (1,1,1) & 0.353 (<0.01) & 0.537 (<0.01)  & 0.362 (<0.01) & 0.662 (<0.01) \\
\midrule
\multirow{7}{*}{CI22} & (1,0,0) & 0.326 (<0.01) & 0.578 (<0.01) & 0.324 (0.02) & 0.899 (<0.01) \\
& (0,1,0) & 0.334 (<0.01) & 0.568 (<0.01) & 0.329 (<0.01) & 1.099 (<0.01) \\
& (0,0,1) & 0.363 (<0.01) & 0.604 (<0.01) & 0.379 (<0.01) & 1.121 (<0.01) \\
& (0,1,1) & 0.332 (<0.01) & 0.472 (<0.01) & 0.328 (<0.01) & 0.816 (<0.01) \\
& (1,0,1) & 0.322 (0.17) & 0.349 (<0.01) & \textbf{0.320 (0.35)} & 0.479 (<0.01) \\
& (1,1,0) & 0.323 (0.17) & 0.495 (<0.01) & \textbf{0.319 (1)} & 0.983 (<0.01) \\
& (1,1,1) & 0.331 (0.18) & 0.402 (<0.01) & 0.333 (0.17) & 0.570 (<0.01) \\
\midrule
\multicolumn{2}{c|}{Benchmark} & \multicolumn{2}{c}{0.429 (<0.01)} & \multicolumn{2}{c}{0.814 (<0.01)} \\
\bottomrule
\end{tabular}}
\caption{Average MSFEs at horizon $h=1$ for GNHAR and JC-GNHAR models. MCS test $p$-values are in brackets. Bold indicates inclusion in the MCS at $20\%$ significance.} 
\label{tab:GNHAR-JCGNHAR-msfe-h1}
\end{table}

\begin{table}[H]
\centering
{\setlength{\tabcolsep}{9pt}\small
\begin{tabular}{ll|cccc}
\toprule
\textbf{Network} & \textbf{Model} & $\boldsymbol{h=5}$ & $\boldsymbol{h=10}$ & $\boldsymbol{h=22}$ & $\boldsymbol{h=44}$ \\
\midrule
\multirow{4}{*}{Fully} & GNHAR(1,0,1) & 0.573 & 0.737 & 1.021 & 1.102 \\
& GNHAR(1,1,0) & 0.559 & 0.784 & 1.066 & 1.146 \\
& JC-GNHAR(1,0,1) & 0.616 & 0.748 & 0.994 & 1.029 \\
& JC-GNHAR(1,1,0) & 0.538 & 0.810 & 1.081 & 1.299 \\
\midrule
\multirow{4}{*}{GC1} & GNHAR(1,0,1) & 0.567 & 0.726 & 0.992 & 1.149 \\
& GNHAR(1,1,0) & 0.566 & 0.772 & 0.952 & 1.113 \\
& JC-GNHAR(1,0,1) & 0.634 & 0.726 & 0.977 & 1.035 \\
& JC-GNHAR(1,1,0) & 0.543 & 0.792 & 1.101 & 1.287 \\
\midrule
\multirow{4}{*}{GC22} & GNHAR(1,0,1) & 0.585 & 0.755 & 0.961 & 1.045 \\
& GNHAR(1,1,0) & 0.572 & 0.744 & 1.110 & 1.181 \\
& JC-GNHAR(1,0,1) & 0.606 & 0.726 & 0.903 & \textbf{0.980} \\
& JC-GNHAR(1,1,0) & 0.566 & 0.783 & 1.126 & 1.171 \\
\midrule
\multirow{4}{*}{CI1} & GNHAR(1,0,1) & 0.571 & 0.851 & 0.865 & 1.060 \\
& GNHAR(1,1,0) & 0.548 & 0.753 & 0.973 & 1.167 \\
& JC-GNHAR(1,0,1) & 0.557 & 0.745 & 0.886 & 1.046 \\
& JC-GNHAR(1,1,0) & 0.534 & 0.787 & 1.020 & 1.109 \\
\midrule
\multirow{4}{*}{CI22} & GNHAR(1,0,1) & 0.556 & \textbf{0.700} & \textbf{0.844} & 1.044 \\
& GNHAR(1,1,0) & 0.550 & 0.721 & 0.874 & 1.111 \\
& JC-GNHAR(1,0,1) & 0.547 & 0.707 & 0.930 & 1.025 \\
& JC-GNHAR(1,1,0) & \textbf{0.530} & 0.709 & 0.861 & 1.062 \\
\midrule
\multicolumn{2}{c|}{HAR Benchmark} & 0.798 & 1.034 & 2.174 & 2.404 \\
\multicolumn{2}{c|}{JC-HAR Benchmark} & 0.953 & 1.289 & 2.402 & 2.563 \\
\bottomrule
\end{tabular}}
\caption{Average MSFEs at horizons $h=5, 10, 22, 44$ for the selected GNHAR and JC-GNHAR models. Bold font highlights the best predictive performance at each horizon.} \label{tab:GNHAR-JCGNHAR-msfe-hlong}
\end{table}

\begin{table}[H]
\centering
{\setlength{\tabcolsep}{7pt}\small
\begin{tabular}{lc|ccccc}
\toprule
\textbf{Model} & \textbf{ARX Order} & $\boldsymbol{h=1}$ & $\boldsymbol{h=5}$ & $\boldsymbol{h=10}$ & $\boldsymbol{h=22}$ & $\boldsymbol{h=44}$ \\
\midrule
HAR Benchmark & 0 & 0.429 & 0.798 & 1.034 & 2.174 & 2.404 \\
JC-HAR Benchmark & 0 & 0.814 & 0.953 & 1.289 & 2.402 & 2.563 \\
\midrule
GNHAR & 0 & 0.322 & 0.556 & \textbf{0.700} & \textbf{0.844} & \textbf{1.044} \\
JC-GNHAR & 0 & 0.320 & \textbf{0.547} & 0.707 & 0.930 & 1.025 \\
\midrule
\multirow{6}{*}{gl-$\lambda$ GNHARX} & 1 & \textbf{0.310} & 0.548 & 0.722 & 0.938 & 1.157 \\
& [0,0,0] & 0.316 & 0.776 & 1.040 & 1.136 & 1.510 \\
& [1,0,0] & 0.326 & 0.692 & 0.928 & 1.458 & 1.682 \\
& [1,0,1] & 0.526 & 0.853 & 1.530 & 2.193 & 2.217 \\
& [1,1,0] & 0.609 & 0.749 & 1.101 & 2.318 & 2.801 \\
& [1,1,1] & 0.907 & 1.163 & 2.784 & 3.176 & 2.671 \\
\midrule
\multirow{6}{*}{gl-$\lambda$ JC-GNHARX} & 1 & 0.311 & 0.570 & 0.758 & 0.984 & 1.167 \\
& [0,0,0] & 0.303 & 0.715 & 1.059 & 1.247 & 1.575 \\
& [1,0,0] & 0.322 & 0.738 & 1.026 & 1.302 & 1.871 \\
& [1,0,1] & 0.435 & 0.860 & 2.115 & 1.568 & 2.011 \\
& [1,1,0] & 0.636 & 0.734 & 1.105 & 2.272 & 2.949 \\
& [1,1,1] & 0.817 & 1.188 & 1.641 & 2.284 & 2.713 \\
\midrule
HARX Benchmark & 1 & 0.591 & 1.435 & 1.713 & 2.345 & 3.396 \\
JC-HARX Benchmark & 1 & 0.635 & 1.604 & 2.136 & 2.829 & 3.758 \\
\bottomrule
\end{tabular}}
\caption{Average MSFEs at horizons $h=1, 5, 10, 22, 44$ for the GNHARX models. Bold font highlights the best predictive performance at each horizon.} \label{tab:GNHARX-msfe}
\end{table}

\end{document}